\documentclass[aps,pre]{revtex4}
\usepackage[pdftex]{graphicx}
\usepackage{color}
\pdfoutput=1 
\def\be{\begin{equation}}
\def\ee{\end{equation}}     
\def\bfi{\begin{figure}}
\def\efi{\end{figure}}
\def\bea{\begin{eqnarray}}
\def\eea{\end{eqnarray}}

\begin{document}

\title{The role of initial state and final quench temperature on the aging properties in phase-ordering kinetics}

\author{Federico Corberi}
\affiliation {Dipartimento di Fisica ``E.~R. Caianiello'', and INFN, Gruppo Collegato di Salerno, and CNISM, Unit\`a di Salerno,Universit\`a  di Salerno, 
via Giovanni Paolo II 132, 84084 Fisciano (SA), Italy.}

\author{Rodrigo Villavicencio-Sanchez}
\affiliation {Dipartimento di Fisica ``E.~R. Caianiello'', Universit\`a  di Salerno, via Giovanni Paolo II 132, 84084 Fisciano (SA), Italy.}

\begin{abstract}

We study numerically the two-dimensional Ising model with
non-conserved dynamics quenched from an initial equilibrium
state at the temperature $T_i\ge T_c$ to a final temperature 
$T_f$ below the critical one. 
By considering processes initiating both from a disordered state at infinite temperature $T_i=\infty$ 
and from the critical configurations at $T_i=T_c$ and spanning the range of
final temperatures $T_f\in [0,T_c[$  we elucidate the role 
played by $T_i$ and $T_f$ on the aging properties and,
in particular, on the behavior of the autocorrelation $C$ and of the 
integrated response function $\chi$.
Our results show that for any choice of $T_f$, while the autocorrelation
function exponent $\lambda _C$ takes a markedly different value for 
$T_i=\infty$ [$\lambda _C(T_i=\infty)\simeq 5/4$] or $T_i=T_c$ [$\lambda _C(T_i=T_c)\simeq 1/8$]
the response function exponents are unchanged. 
Supported by the outcome of the analytical solution of the solvable
spherical model we interpret this 
fact as due to the different contributions provided to autocorrelation and
response by the large-scale properties of the system. 
As changing $T_f$ is considered, although this is expected to play no role 
in the large-scale/long-time properties of the system, we show important
effects on the quantitative behavior of $\chi$. In particular, data for 
quenches to $T_f=0$ are consistent 
with a value of the response function exponent $\lambda _\chi=\frac{1}{2}\lambda _C(T_i=\infty)=5/8$ different from 
the one [$\lambda _\chi \in (0.5-0.56)$] found in a wealth of previous numerical
determinations in quenches to finite final temperatures.
This is interpreted as due to important pre-asymptotic corrections 
associated to $T_f>0$.

\end{abstract}

\maketitle

\section{Introduction}

Slow evolution is usually observed when glassy and disordered materials or binary systems are 
quenched across a phase transition \cite{cuglia,BCKM,ioeleti}.  However, while the additional
difficulties related to glassiness and disorder make understanding the formers quite difficult,
the phase-ordering process occurring when a clean magnet is cooled below the critical
point is a simpler context where aging properties can more easily be investigated.
In particular, an important issue concerns
the scaling properties of two-times quantities such as the autocorrelation $C$ and the 
response function $\chi$. The mutual relation between them is of paramount importance 
for glassy systems because, when suitable conditions are met, it encodes the 
still debated structure of the equilibrium states via the Franz-Mezard-Parisi-Peliti theorem \cite{fmpp}.  A program towards a full understanding of this relationship might start
very naturally from the paradigmatic case of non-disordered magnets. 
Despite their relative simplicity, however, a fully satisfactory 
reference analytical theory of aging in these systems does not exist and their behavior
is not fully understood. 

In this paper we present a rather complete numerical study of the dynamics of the possible 
sub-critical quenches of a bi-dimensional magnet described by the Ising model with
spin-flip dynamics. In particular, we study systems cooled from an initial equilibrium state
at an infinite temperature $T_i=\infty$ and from the critical configuration at the 
transition temperature $T_i=T_c$. Starting from these initial states, we consider deep quenches to a final temperature $T_f=0$, or to an intermediate
one $T_f=0.66 \,T_c$, and shallow coolings to $T_f=0.97 \,T_c$.
This allows us to discuss the behavior of the response function in the processes ending at
$T_f=0$ (starting with arbitrary $T_i$) and in those starting from $T_i=T_c$ 
(ending at any $T_f$), which were never studied before and yield unexpected results. 

As expected on the basis of dynamical scaling \cite{bray,furukawa}, we find that observable
quantities take scaling forms regulated by universal exponents.  
Our data for quenches to $T_f=0$ show quite unambiguously that, in this case,
the response function exponent $\lambda _\chi$ [see Sec. \ref{Model}, 
Eqs. (\ref{scalChi},\ref{largeh}), for a definition of these quantities]
takes a value compatible with 
$\lambda _\chi =\frac{1}{2}\lambda _C\simeq 5/8$, $\lambda _C $ being the exponent
governing the decay of the autocorrelation function 
[see Eqs. (\ref{scalC},\ref{largec})].
This determination of $\lambda _\chi$ is definitely different from the one found 
in previous studies \cite{unquarto,algo,henkelrough} which were focused on quenches to final $T_f$. 
Since the final temperature 
of the quench is 
expected to be an irrelevant parameter \cite{bray,brayren}, in the sense of the renormalization group, we conjecture
that $\lambda _\chi=\frac{1}{2}\lambda  _C$ is the correct asymptotic value and that, in 
the above mentioned previous studies, the correct value was shadowed by finite-$T_f$ 
preasymptotic corrections (which are indeed observed also in the present study when $T_f$
is chosen finite). 

Concerning the role of the initial condition we show that, while the autocorrelation function takes
a radically different value \cite{brayfromcrit} when the quench is made from 
$T_i=\infty$ [$\lambda _C(T_i=\infty)=5/4$] or $T_i=T_c$ [$\lambda _C(T_i=T_c)\simeq 1/8$],
the response function exponent remains basically unchanged. By solving the 
large-$N$ (or spherical) model we find that the same feature is shared also in this
analytically tractable model of magnetism. Physically, we interpret the different
sensitivity of  $C$ and $\chi$ to the initial state as due to the different role played by the 
large-scale properties of the system which, in the quench from $T_i=T_c$, keep 
memory of the critically correlated initial state.

This Article is organized as follows: In Sec. \ref{Model} we set the notation, define the 
model under investigation, the different observables considered throughout the paper
and their scaling behavior.
Sec. \ref{SIMUL} is devoted to the presentation and discussion of our numerical results.
In Sec. \ref{largen}, by exactly solving the large-$N$ model, we discuss
some analogies with the numerical results presented in Sec. \ref{SIMUL}.
In Sec. \ref{concl} we summarize the main findings of this paper, discuss 
some open issues, and present the conclusions of the work.

\section{The Ising Model and the observable quantities} 
\label{Model}

We consider a system of ${\cal N}$ spins on a lattice with the Ising Hamiltonian
\be
H=-J \sum_{<ij>}\sigma_i \sigma_j
\ee
where the sum runs over the nearest neighbors pairs $<ij>$ and $J>0$. 
The time evolution occurs through single spin-flip dynamics with Glauber
transition rates \cite{glauber}
\be
w_i([\sigma] \to [\sigma'])=\frac{ 1}{2}
\left[ 1-\sigma_i \tanh\left(\frac{h_i^W}{T}\right)\right]
\label{w}
\ee
where $[\sigma]$ and $[\sigma']$ are spin configurations differing
only for the value of the spin on the $i$-th site and
$h_i^W=J\sum_{k\in \{nn_i\}} \sigma_k$, where the sum is restricted to the nearest 
neighbors $\{nn_i\}$ of $i$, is the local Weiss field.

In a quenching protocol the system is prepared at $t=0$ 
in an equilibrium configuration at the initial temperature $T_i$ and 
is then evolved with the transition rates (\ref{w}) where $T$ is set equal to
the final temperature $T_f$.

The typical size of the growing ordered domains $L(t)$ will be computed in this
work as the inverse excess energy:
\be
L(t)=[E(t)-E_\infty]^{-1} .
\label{lt}
\ee
Here, $E(t)=\langle H(t)\rangle$ is the average energy at time $t$, and $E_\infty $ is the one of the 
equilibrium state at the final temperature $T_f$. Here and in the following the average $\langle 
\cdots \rangle$ is taken over different realizations of the initial state and of the thermal histories,
namely over the random flip events generated by the transition rates (\ref{w}).  
Equation~(\ref{lt}) is often used to determine $L(t)$ \cite{bray} because the excess energy of the coarsening system with respect to the equilibrated one is associated to the density of domain walls which,
in turn, is inversely proportional to the typical domain size.  

The two-points/two-times correlation function is defined as
\be
{\cal G}(\vec r,t,s)=\langle \sigma _i (t) \sigma _j (s)\rangle - \langle \sigma _i (t)\rangle
\langle \sigma _j (s)\rangle,
\label{defcalG}
\ee
where we assume $t\ge s$,
which depends only on 
the distance $\vec r$ between sites $i$ and $j$ due to space homogeneity.
In the phase ordering process this quantity takes the additive structure
\be
{\cal G}(\vec r,t,s)={\cal G}_{st}(\vec r,t-s)+{\cal G}_{ag}(\vec r,t,s),
\label{calG}
\ee
where the first contribution describes the fast equilibrium fluctuations in the pure states 
which are attained well inside the domains and the second one contains the non-equilibrium aging properties.
Being an equilibrium contribution, the first term in Eq. (\ref{calG}) vanishes over distances larger than the equilibrium coherence length $\xi _{eq}$ and/or time differences longer than the 
equilibrium correlation time $\tau _{eq}$. In the present paper we will be interested in the 
behavior of the system on distances and time-differences much larger than $\xi _{eq}$ 
and $\tau _{eq}$, respectively, where the first term of Eq. (\ref{calG}) can be neglected.
We will then focus on the second contribution, the aging term, and drop the suffix $_{ag}$.

Letting $t=s$ in Eq. (\ref{defcalG}) amounts to consider the equal-time correlation function 
\be
G(\vec r,t)=\langle \sigma _i (t) \sigma _j (t)\rangle,
\label{defG}
\ee
where we have neglected the last term in Eq. (\ref{defcalG}) since in the processes
we will be interested in one always has $\langle \sigma _i(t)\rangle=0$.
We compute numerically this quantity as
\be
G(r,t)=\frac{1}{{4\cal N}} \sum _{i,j:|i-j|=r} \left \langle 
\sigma _i (t) \sigma _j (t)\right \rangle.
\ee
where the sum runs over all the $4{\cal N}$ couple of sites at distance $r$ on the horizontal
and vertical direction.
This correlation function obeys the scaling form \cite{bray,furukawa}
\be
G(r,t)=g\left [\frac{r}{L(t)}\right ],
\label{scalG}
\ee
where $g(y)$ is a scaling function. For completeness, let us mention that 
corrections to the form (\ref{scalG}) were reported in \cite{epl}. These corrections,
however, are negligible for the large system-sizes considered in this paper. 
The sharp nature of the domains walls imply
a short-distance behavior \cite{bray,porod}
\be
g(y)\simeq 1-ay,
\label{porod}
\ee
where $a$ is a constant, in the limit $y\ll 1$.

When the system is quenched from an initially critical state, i.e. with $T_i=T_c$,
the scaling function has the following large-$y$ behavior \cite{brayfromcrit}
\be
g(y)\propto y^{-(d-2+\eta)},
\label{critg}
\ee
where $\eta$ is the equilibrium critical exponent, namely $\eta =1/4$ in the two-dimensional case 
studied in this paper.
This simply expresses the fact that, for distances $r\gg L(t)$ much larger than those where
ordering has been effective at the current time, the system keeps memory of  the equilibrium 
critical initial state. As we will see shortly, this bears important consequences, particularly on the
behavior of the autocorrelation function $C$ that is defined by
setting $i=j$ in Eq. (\ref{defcalG})
\be
C(t,s)=\langle \sigma _i (t) \sigma _i (s)\rangle.
\label{defC}
\ee
This quantity does not depend on $i$ due to homogeneity.
Enforcing this, $C$ will be computed as 
\be
C(t,s)=\frac{1}{{\cal N}}\sum _i \left < 
\sigma _i (t) \sigma _i (t)\right \rangle,
\label{spataveC}
\ee
in our simulations.
The autocorrelation obeys the scaling form \cite{bray,furukawa}
\be
C(t,s)=c\left [\frac{L(t)}{L(s)}\right ],
\label{scalC}
\ee
where $c(x)$ is a scaling function with the large-$x$ behavior
\be
c(x)\simeq x^{-\lambda _C}.
\label{largec}
\ee
The autocorrelation exponent is expected to be $\lambda _C(T_i=\infty)=5/4$ \cite{bray,liumaz}
for a quench from $T_i=\infty$ and a much smaller value \cite{brayfromcrit}
$\lambda _C(T_i=T_c)=1/8$ for a quench from the critical state at $T_i=T_c$.

The linear response function is defined as
\be
R(t,t')=\left . \frac {\delta \langle \sigma _i (t) \rangle}{\delta h_i (t')} \right \vert _{h=0},
\label{defR}
\ee
where $h_i(t')$ -- the perturbation -- is a magnetic field of amplitude $h$ applied on site 
$i$ only at time $t'$. From this quantity the integrated response function,
referred to also as
dynamic susceptibility or zero-field-cooled susceptibility, 
is obtained as 
\be
\chi (t,s)=\int _s^t R(t,t')\,dt'.
\label{defChi}
\ee
Given that also the response function is independent on $i$,
we improve the numerical efficiency by computing this quantity as a spatial 
average, similarly to what done for the
autocorrelation function [Eq. (\ref{spataveC})].
We obtain this quantity using the generalization to non-equilibrium states
of the fluctuation-dissipation theorem derived in \cite{algo,algo2}. Similarly to the
well known equilibrium theorem, this amounts to an analytical relation  
between $\chi $ and certain correlation functions of the {\it unperturbed} system, 
namely the one where the perturbation $h$ is absent. The great advantage of this approach is
the fact that the $h\to 0$ limit in Eq. (\ref{defR}) is dealt with analytically, making the numerical computation of the response function totally reliable and very efficient. 
Notice that the use of the non-equilibrium fluctuation-dissipation relation provides
directly the quantity $T_f\chi$.

Being related to correlation functions also the response function can be split in two
terms, similarly to Eq. (\ref{calG}). However, at variance with the other quantities
discussed above, the term $\chi _{st}$ is not negligible and, in order to isolate the
aging part, it has to be subtracted away. This can be done
by computing preliminarily $\chi _{st}$ in the equilibrium state, as described in \cite{unquarto}.

The aging part of the response function obeys the scaling form \cite{revresp}
\be
\chi(t,s)=L(s)^{-\alpha}h\left [\frac{L(t)}{L(s)}\right ],
\label{scalChi}
\ee
where $h(x)$ is a scaling function with the large-$x$ behavior
\be
h(x)\simeq x^{-\lambda _\chi}.
\label{largeh}
\ee
It is expected that $\chi$ becomes independent on $s$ for large time differences
$t-s\gg s$. Because of Eqs. (\ref{scalChi},\ref{largeh}) this implies that 
\be
\alpha=\lambda _\chi. 
\label{lameqalf}
\ee
This property has been verified several times in the 
literature \cite{unquarto}. 
As for the value of the response exponent $\alpha$
it was conjectured to be $\alpha =1/2$ and  
present numerical determinations set its value in the range $[0.5-0.56]$.
We will discuss further the value of this exponent in the following.

\section{Numerical simulations} \label{SIMUL}

In this Section we present and discuss the results of our simulations
which have been obtained by quenching a two-dimensional system of linear size 
$\Lambda =2\cdot 10^3$ (unless differently specified)
with $J=1$ and periodic boundary conditions. 
We have evolved the system starting with configurations from the 
infinite as well as the critical temperature
from $t=0$ to a final time $t=3\cdot 10^4$.
In the case of quenches from the 
critical state, in order to equilibrate the system at $T_c$, we have used
the Wolff algorithm \cite{wolff}. Data are organized in different sections
according to the different choices of the initial and final temperatures
of the various quenching protocols considered.

\subsection{Quenches to $T_f=0$} \label{to0}

\subsubsection{From $T_i=\infty$} \label{TinfT0}

\begin{figure}
\rotatebox{0}{\resizebox{.95\textwidth}{!}{\includegraphics{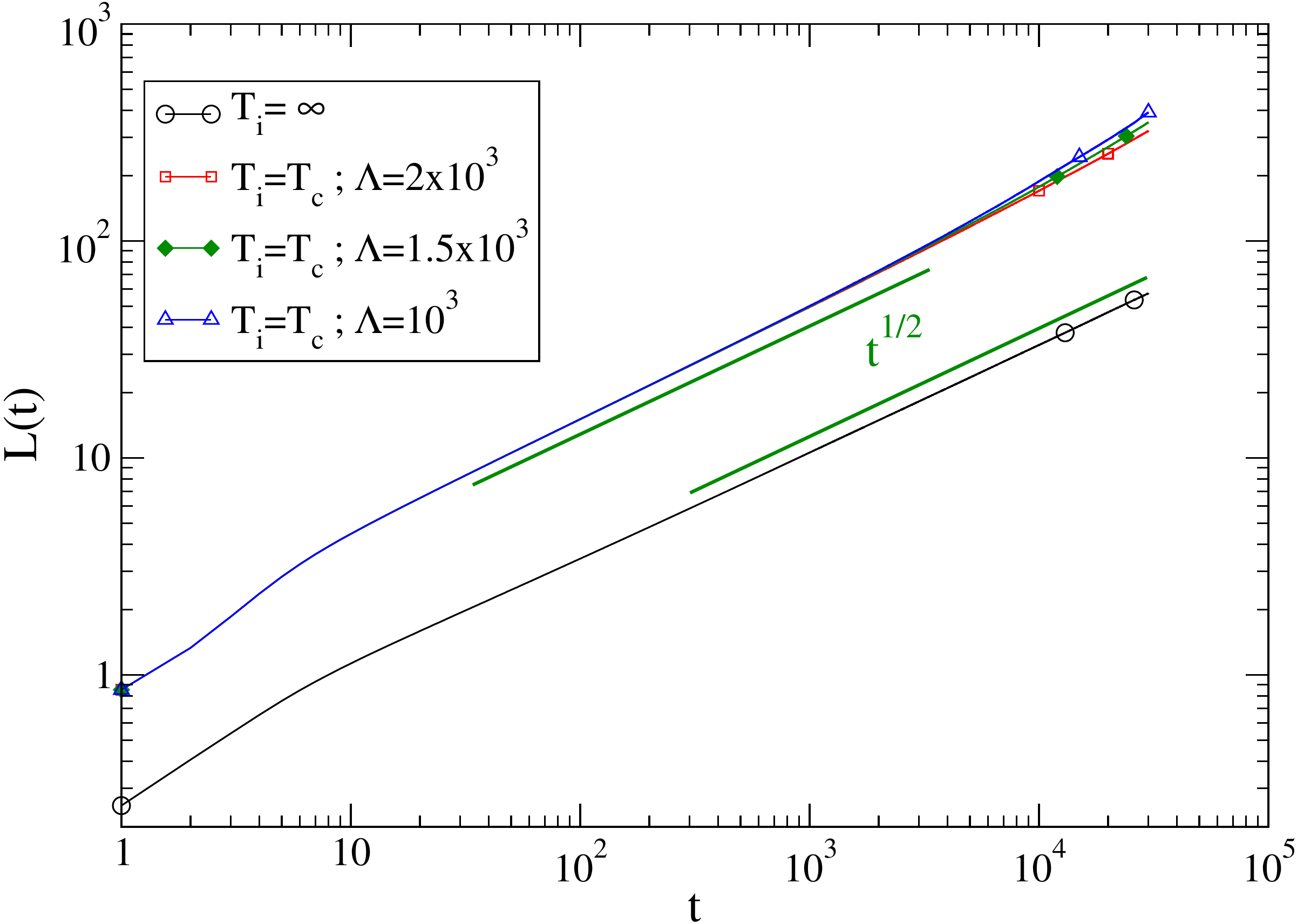}}}
\caption{(Color on-line). $L(t)$ is plotted against $t$ on a log-log plot
for the quench of
the system to $T_f=0$ starting from the equilibrium state at $T_i=\infty$
(lower black curve with a circle)
and at $T_i=T_c$ (upper partially collapsing three curves, corresponding to three different
system sizes $\Lambda $ as detailed in the key). The bold green lines are the behavior
$L(t)\sim t^{1/2}$ expected for an infinite system for large times.}
\label{L_to_zero}
\end{figure}

The behavior of $L(t)$ is shown in Fig. \ref{L_to_zero} (lower black curve).
Starting from $t\simeq 10$, the expected power-law $L(t)\sim t^{1/z}$
with $z=2$ sets in (a fit in the last decade provides $z=2.004$).
Data for smaller system sizes ($\Lambda =1.5\cdot 10^3$ and $\Lambda =10^3$, not shown) 
superimpose to the plotted data. This, and the power-law behavior of $L$, clearly indicate
that our simulations are finite-size effect free in the present case.

\begin{figure}
\rotatebox{0}{\resizebox{.95\textwidth}{!}{\includegraphics{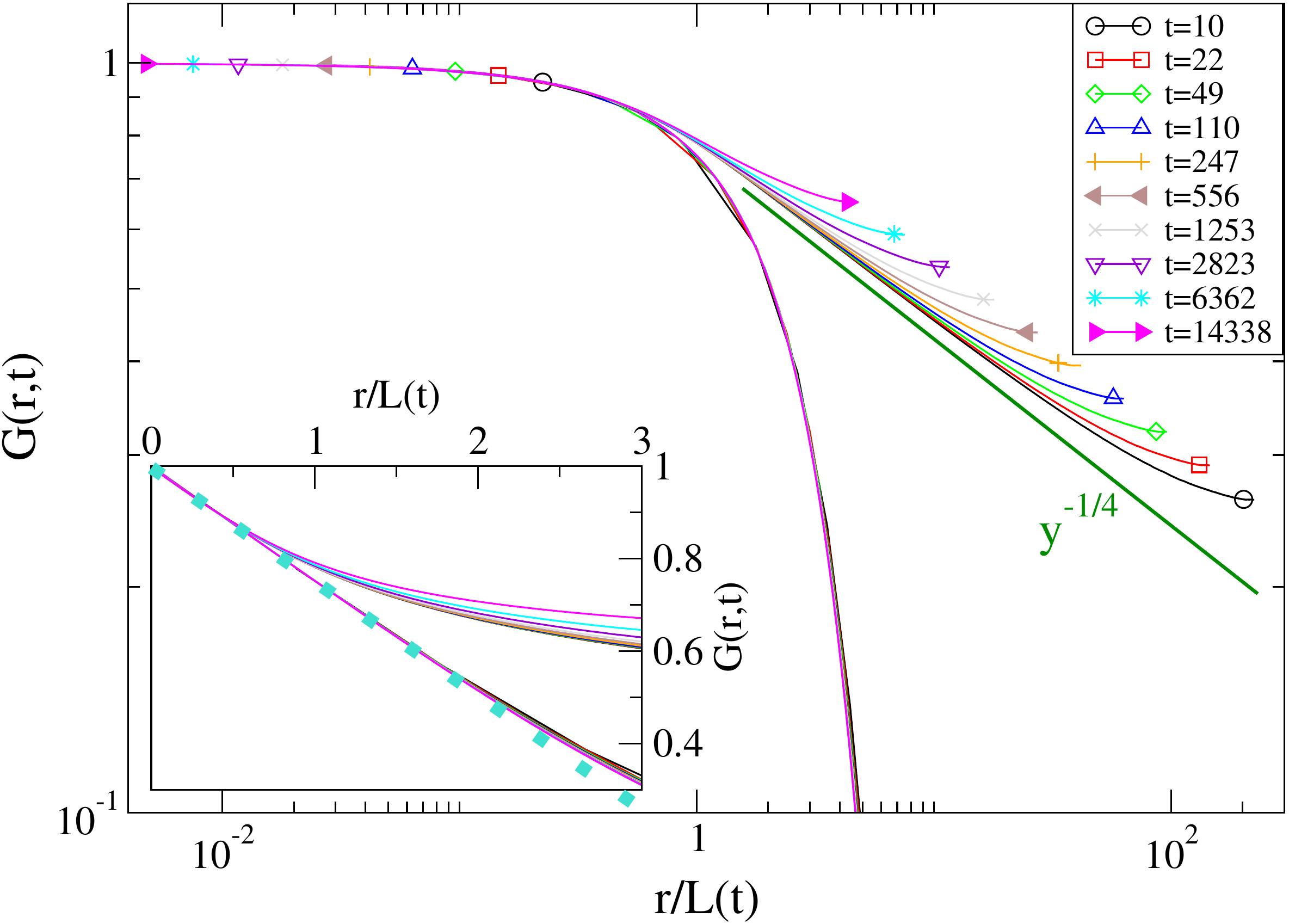}}}
\caption{(Color on-line). The correlation function $G(r,t)$ for a quench to $T_f=0$ 
is plotted against $y=r/L(t)$ 
on a double-logarithmic plot at the different times indicated in the key
(these are determined as to be exponentially spaced by an automatic routine). 
The lower set of collapsing curves (without symbols), correspond to the case 
discussed in Sec. \ref{TinfT0} of a quench from an equilibrium state at the initial
temperature  $T_i=\infty$. The upper group of curves,
marked with a symbol, correspond to the case discussed in Sec. \ref{TcT0} of  a quench
starting from the critical state, i.e. $T_i=T_c$.
The bold green straight line is the power-law $y^{-1/4}$ of Eq. (\ref{critg}).
In the inset a zoom of the same data in the region of small $r/L(t)$ is plotted on a 
double-linear plot. The bold-dotted turquoise line is the linear behavior 
of Eq. (\ref{porod}), namely the Porod's law.}
\label{G_to_zero}
\end{figure}

The behavior of the equal-time correlation function is shown in 
Fig. \ref{G_to_zero} (lower set of curves) for various times (see key)
on a double-logarithmic plot.
As expected on the basis of Eq. (\ref{scalG}), an excellent data collapse of the
curves at different times is obtained 
by plotting $G$ against the rescaled space $y=r/L(t)$ in all the region where 
$G$ is significantly larger than zero.  As it can be seen in the inset of Fig. \ref{G_to_zero},
where a zoom on the small-$r$ sector is presented on a plot with linear axis, 
a linear behavior -- the so called Porod's law, Eq. (\ref{porod}) -- is well obeyed in this regime,
as discussed in Sec. \ref{Model}.

\begin{figure}
\rotatebox{0}{\resizebox{.85\textwidth}{!}{\includegraphics{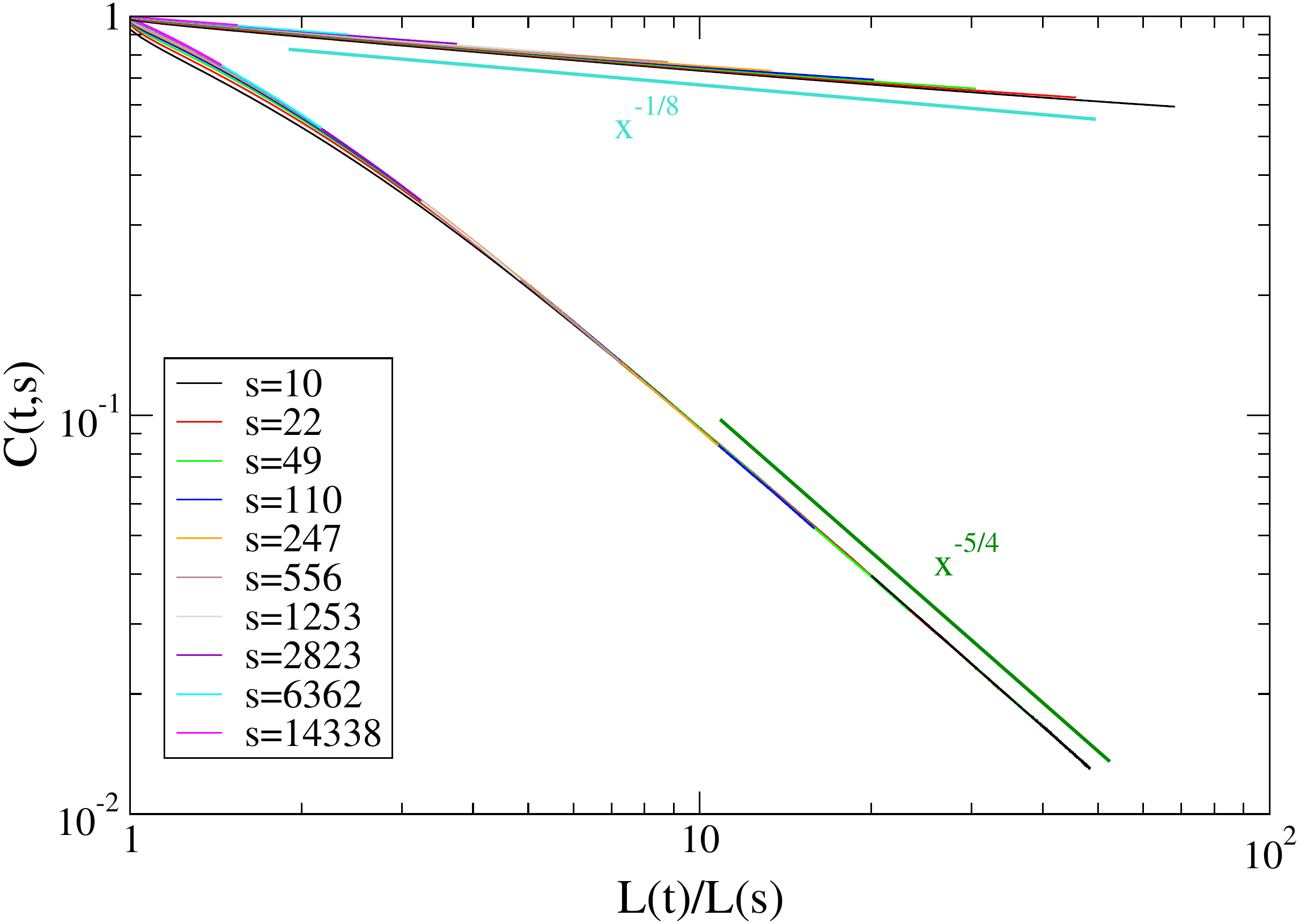}}}
\caption{(Color on-line). $C(t,s)$ is plotted against $x=L(t)/L(s)$ 
for the quench of
the system to $T_f=0$ starting from the equilibrium state at $T_i=\infty$
(lower set of curves)
and at $T_i=T_c$ (upper group of curves) for different times (see key).
The bold dark-green and turquoise lines are the expected 
power-laws $x^{-\lambda _C}$ with $\lambda _C=5/4$ and $\lambda _C=1/8$
for the quenches from $T_i=\infty$ and from $T_i=T_c$, respectively.}
\label{C_to_zero}
\end{figure}

The autocorrelation function is plotted against $x=L(t)/L(s)$
in Fig. \ref{C_to_zero}.
The collapse expected on the basis of Eq. (\ref{scalC}) is excellent in the region of large $x$.
For small values of $x$ the superposition is worse for the smaller
values of $s$ but, also in this region, an excellent
collapse is recovered for sufficiently large values of $s$.
This can be interpreted as due to the presence of preasymptotic corrections to scaling
which are not completely negligible for the smaller values of $s$ considered in our simulations.
Notice that for large $x$ the expected behavior $C(x)\sim x^{-\lambda_C}$ of 
Eq. (\ref{largec}) with
$\lambda _C=5/4$ is very well
reproduced (a fit of the curve with $s=10$ for $x\ge 30$ gives $\lambda _C=1.256$).

The data presented insofar show that an excellent scaling behavior is displayed starting from 
the region of moderate times. Therefore this case is an optimal playground to assess the scaling properties
of the response function, which in the past have been the subject of some controversies.

\begin{figure}
\rotatebox{0}{\resizebox{.85\textwidth}{!}{\includegraphics{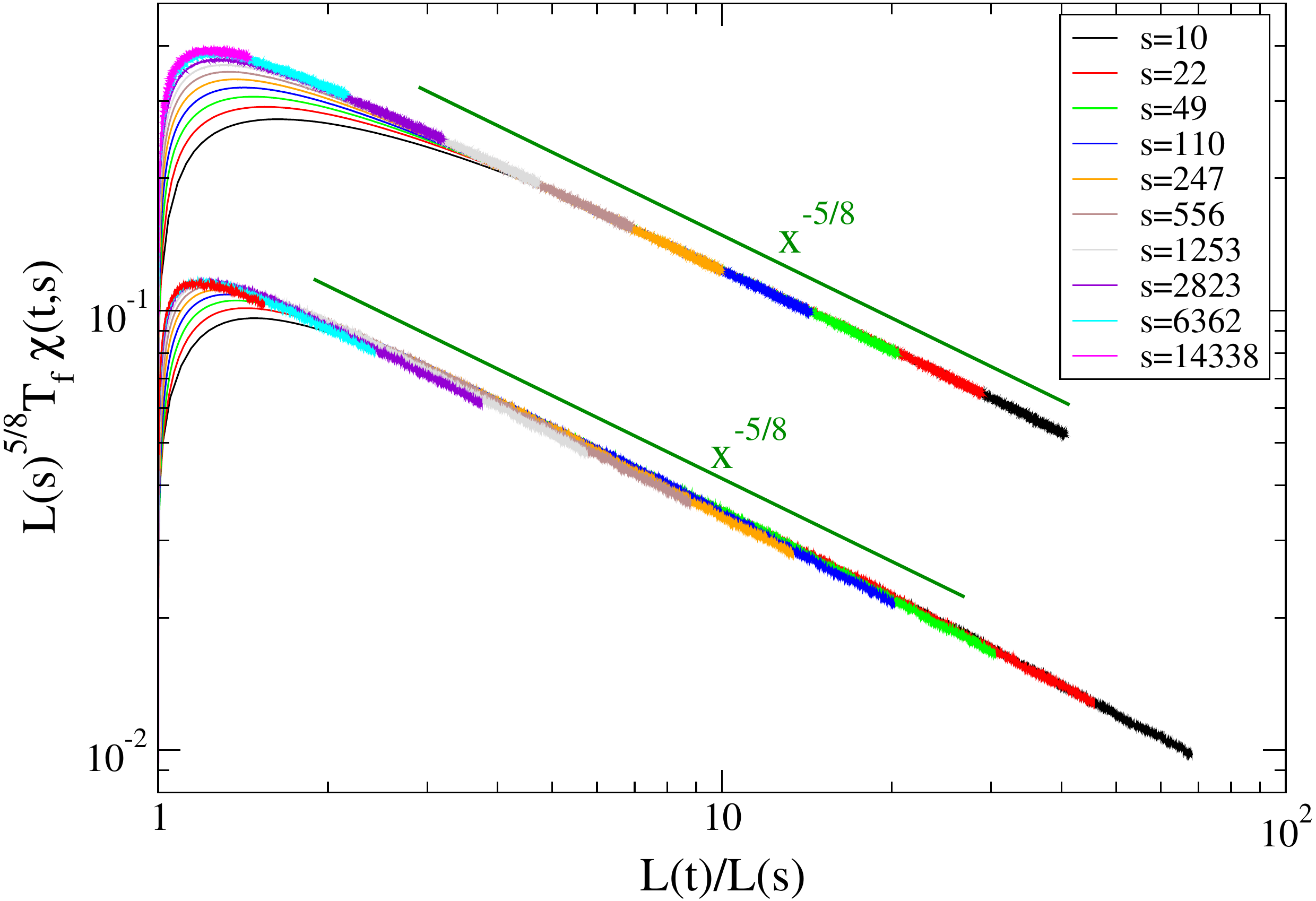}}}
\caption{(Color on-line). $L(s)^{5/8}T_f\,\chi(t,s)$ is plotted against $x=L(t)/L(s)$ 
for the quench of
the system to $T_f=0$ starting from the equilibrium state at $T_i=\infty$
(upper set of curves)
and at $T_i=T_c$ (lower group of curves) for different times (see key).
The bold dark-green line is the 
power-law $x^{-\lambda _\chi}$ with $\lambda _\chi=5/8$.}
\label{Chi_to_zero}
\end{figure}

The response function is shown in Fig. \ref{Chi_to_zero}.
According to Eq. (\ref{scalChi}) and the discussion thereafter one should get 
the collapse of curves for different waiting times $s$ 
by plotting $L(s)^{\alpha}\chi (t,s)$ against $x=L(t)/L(s)$,
where the exponent $\alpha$ equals the exponent $\lambda _\chi$
regulating the large-$x$ behavior of $\chi$ according to Eq. (\ref{largeh}).
Comparison of the large-$x$ behavior of the curves in Fig. \ref{Chi_to_zero}
with the dark-green line $x^{-5/8}$ shows that $\chi$ is 
very well compatible with a value $\lambda _\chi=5/8$.
This value is different from the one found for quenches to finite temperatures
where a value in the range $\alpha \simeq 0.5-0.56$ (according to different determinations)
was reported. This was interpreted \cite{noirough,henkelrough} as the value $\alpha =1/2$ expected on the basis of 
an argument associating the properties of the response to the roughening of the interfaces.
The present determination $\lambda_\chi \simeq 5/8$, instead, suggests that the somewhat 
different value $\alpha=\lambda_\chi=(1/2)\lambda _C$ could be the asymptotically correct one.
Notice that a value $\alpha \simeq 0.6$, roughly comparable to the one we find here,
was found in \cite{RBIM} in the phase-ordering of a weakly disordered magnet in the
limit of an extremely deep quench.

\begin{figure}
\rotatebox{0}{\resizebox{.85\textwidth}{!}{\includegraphics{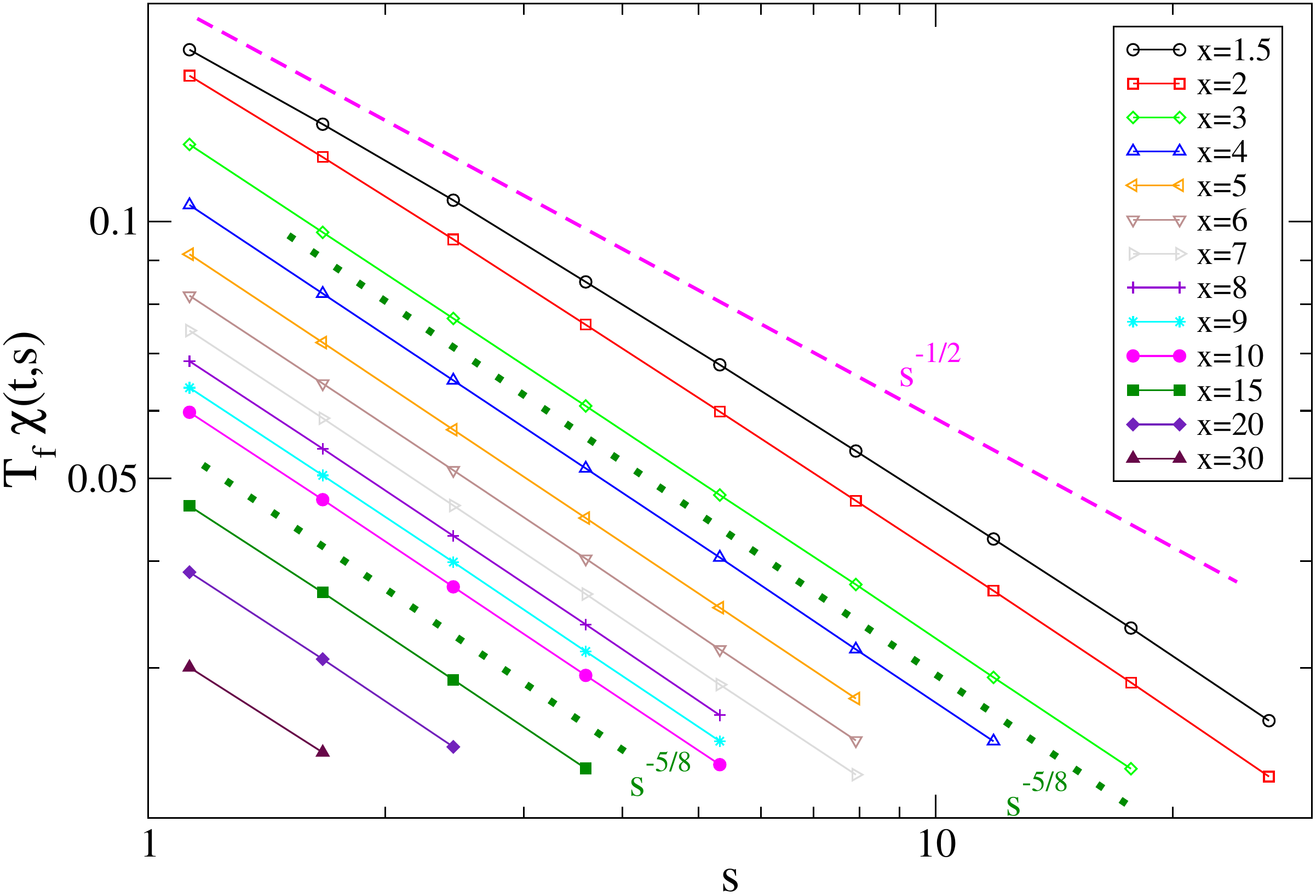}}}
\caption{(Color on-line). $T_f\,\chi(t,s)$ is plotted against $L(s)$ for fixed values of $x$
(indicated in the key) for the quench of
the system to $T_f=0$ starting from the equilibrium state at $T_i=\infty$.
The bold dotted dark-green lines and the dashed magenta one are the 
power-laws $x^{-\alpha}$ with $\alpha=5/8$ and $\alpha=1/2$, respectively.}
\label{scal_vs_tw_fix_x}
\end{figure}

In order to verify better this conjecture and to have the most 
reliable determination of $\alpha$ from our data we plot in Fig. \ref{scal_vs_tw_fix_x}
$T_f\chi (t,s)$ against $L(s)$ for fixed values of $x$ spanning the entire sector $x\in [1.5,30]$.
According to Eq. (\ref{scalChi}) on a double logarithmic scale the slope of the curves
in this figure directly provides the exponent $\alpha $. 
The data of Fig. \ref{scal_vs_tw_fix_x} show consistency with an exponent $\alpha =5/8$ 
(dark-green bold dotted line), while $\alpha=1/2$ does not fit equally well (except, perhaps,
in a region of very small $s$ and $x$ where pre-asymptotic effects may show up,
see discussion below).
Fitting the curves for the different values of $x$, indeed, gives effective exponents 
$\alpha _{eff}=0.575,\, 0.600,\, 0.612,\, 0.615,\, 0.615,\, 0.617,\, 0.616,\, 0.616,\, 0.616,\, 
0.614,\, 0.612,\, 0.611,\, 0.589$ which (with the exception of the first two and of the last 
value, see discussion below) are rather close to $\lambda _C/2=0.625$, while they 
seem to rule out the possibility $\alpha= 1/2$. 

Let us comment shortly on the somewhat smaller values of $a_{eff}$ found
for small $x$ ($x\le 2$) (namely $\alpha _{eff}=0.575$ for $x=1.5$ and 
$\alpha _{eff}=0.600$ for $x=2$)
and for very large $x$ (namely $\alpha _{eff}=0.589$ for $x=30$).
Data for small $x$ are probably affected by pre-asymptotic
effects at small $s$. Indeed, fitting for instance the curve relative to $x=1.5$  
only for values of $L(s)>10$ the effective exponent rises to $a_{eff}=0.609$ 
(from the value $a_{eff}=0575$ when fitted over the whole range of $L(s)$).
The data for $x$ as large as $x=30$ contain only two points and this makes 
the determination of $a_{eff}$ in this case probably insecure.
From this analysis we can conclude therefore that, except in the region of very
small or very large $x$ where preasymptotic corrections and other effects make the 
determination of the exponent insecure, a value $\alpha =5/8$ is well consistent
with the data.

\begin{figure}
\vspace{1cm}
\rotatebox{0}{\resizebox{.45\textwidth}{!}{\includegraphics{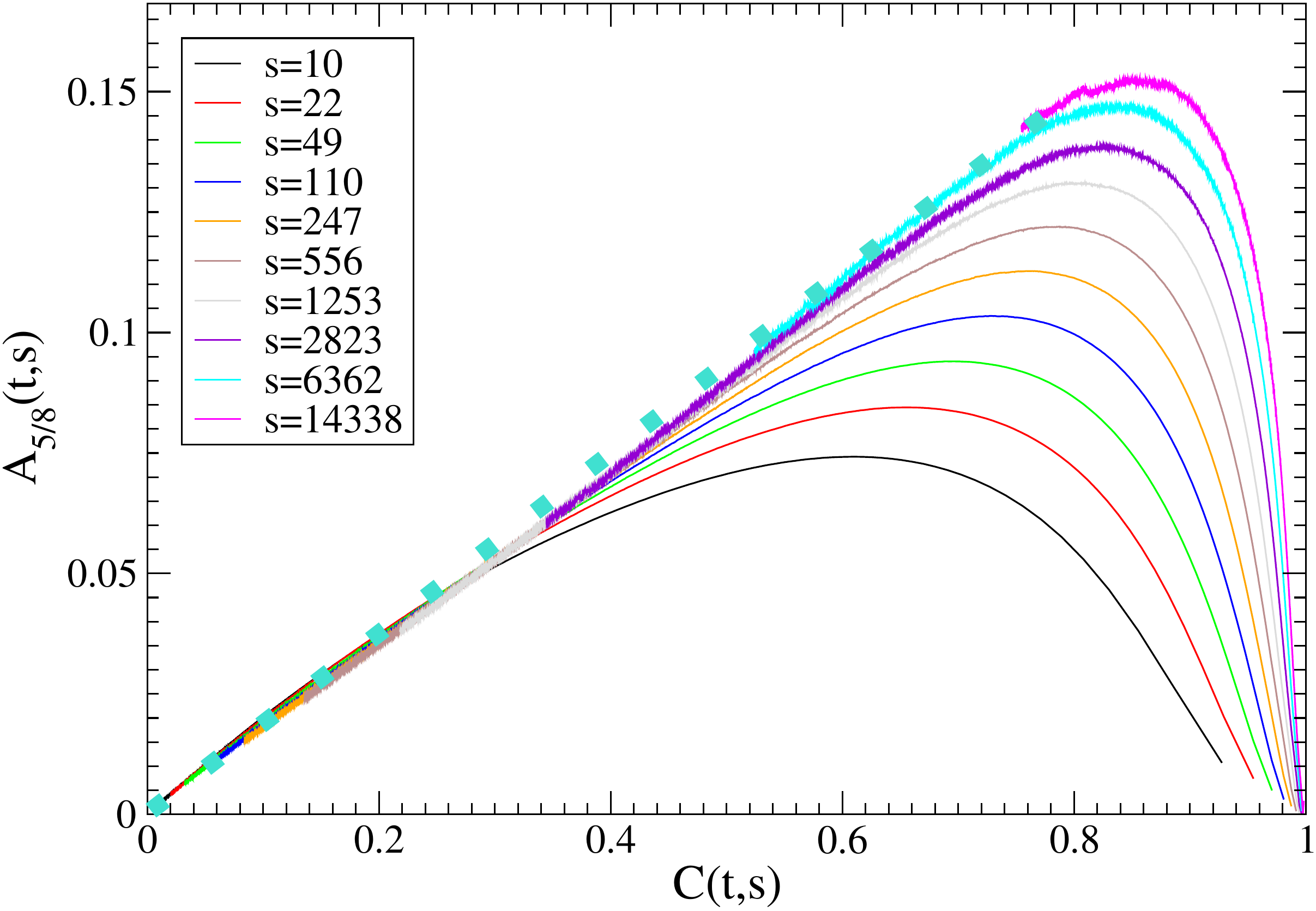}}}
\rotatebox{0}{\resizebox{.45\textwidth}{!}{\includegraphics{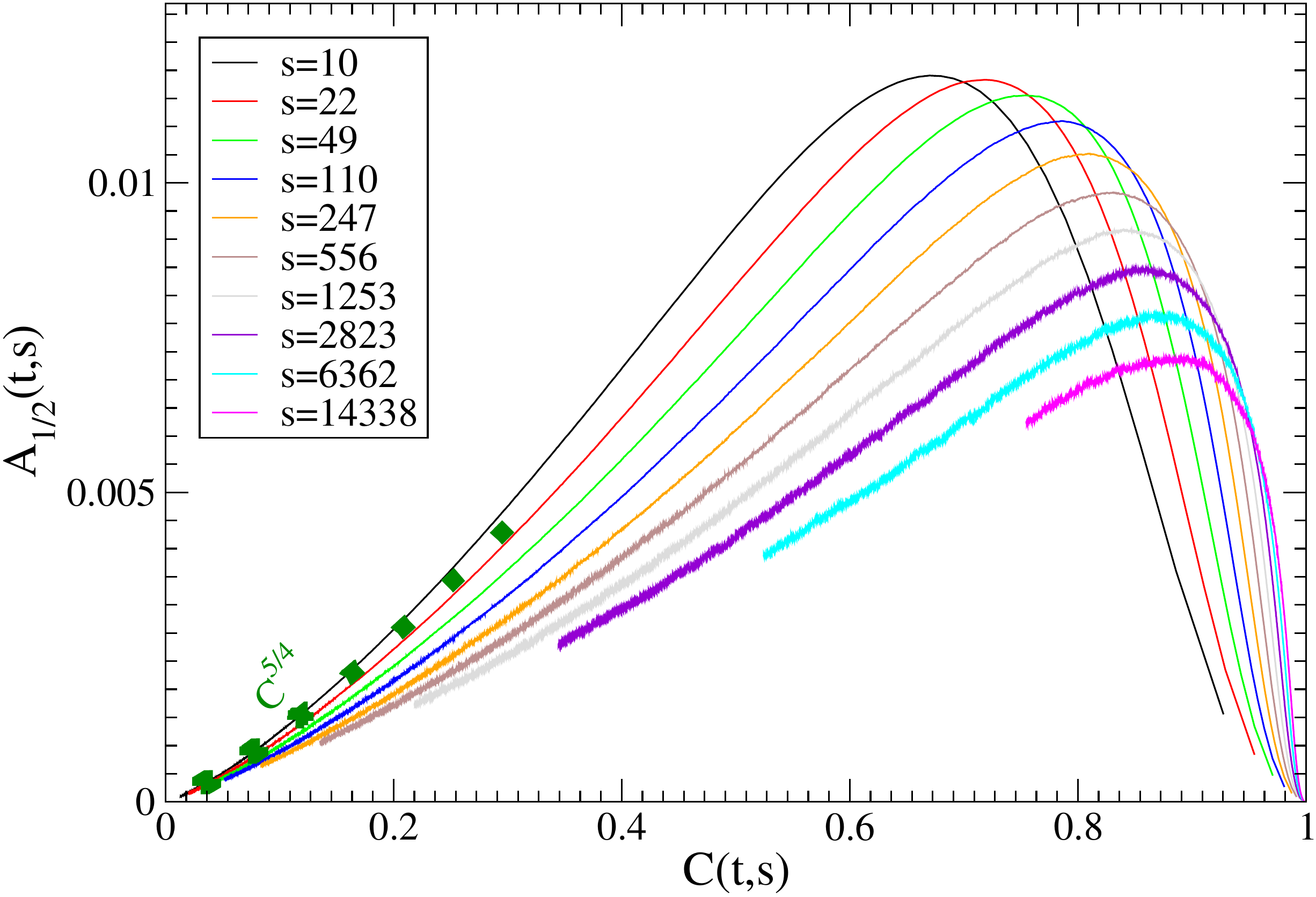}}}
\caption{(Color on-line). The function $A_{5/8}(t,s)$ (left panel)
and $A_{1/2}(t,s)$ (right panel)
are plotted against $C(t,s)$ for the quench of
the system to $T_f=0$ starting from the equilibrium state at $T_i=\infty$
for different times (see key).
The bold-dotted turquoise line on the left panel is guideline for the linear behavior
$[L(s)^{5/8}\chi (t,s)]^2\propto C(t,s)$. The bold-dotted green one on the right panel 
is the behavior $C(t,s)^{5/4}$.}
\label{Chi2_to_zero}
\end{figure}

To provide the most possible reliable determination of the response function exponent
we present a further, alternative analysis of the data in the following.  
Let us observe that, from the scaling of $C$ and $\chi$, Eqs. (\ref{scalC},\ref{scalChi}),
in the region of large time differences where the forms of Eqs. (\ref{largec},\ref{largeh})
hold, one has
\be
A_\alpha(t,s)\equiv \left [s^{\alpha}\, T_f\,\chi (t,s) \right ]^{\frac{\lambda _C}{\alpha}}=\kappa \,C(t,s),
\hspace{1cm} t-s\gg s,
\label{check}
\ee
where we have used the fact that $\alpha =\lambda _\chi$ and $\kappa $ is a constant.
Eq. (\ref{check}) represents a tool for a stringent test on the value of the exponent 
$\alpha $, as we discuss below. In Fig. \ref{Chi2_to_zero} we show the parametric plot of 
$A_\alpha(t,s)$ against 
$C(t,s)$. 
Specifically, for any couple of times $t,s$ we plot $A_\alpha(t,s)$ 
on the vertical axis and $C(t,s)$ on the horizontal one. 
Notice that one can re-parametrize only one (say $t$) of the two-times
appearing in $C(t,s)$ through the value of $C$ itself. In doing that one goes from
$A_\alpha (t,s)$ to a new function $A_\alpha (C,s)$ that, in principle, depends also on $s$
(besides the quantity $C$ on the horizontal axis).
However, according to Eq. (\ref{check}), for large time-differences $t-s\gg s$, meaning small values
of the quantity $C$,  the quantity $A_\alpha $ ought to be an $s$-independent linear function of $C$
(i.e. curves for different values of $s$ should collapse), if the value of $\alpha $ is the 
appropriate one. On the other hand, for an improper value $\beta\ne \alpha$ of this
exponent, instead of Eq. (\ref{check}) the function $A_\beta$ would behave as
\be
A_{\beta}(t,s)\equiv \left [s^{\beta}\, T_f\,\chi (t,s) \right ]^{\frac{\lambda _C}
{\beta}}=\widetilde \kappa s^{\left (1-\frac{\alpha}{\beta}\right )\lambda _c}\,
C(t,s)^{\frac{\alpha}{\beta}},
\hspace{1cm} t-s\gg s,
\label{anticheck}
\ee
where $\widetilde \kappa=\kappa ^{\frac{\alpha}{\beta}}$ is another constant.
Therefore in this case the parametric plot of $A_\beta$ against $C$ would not be linear and
curves for different values of $s$ would not collapse. This qualifies 
this kind of plot as a strict
check on the correct value of $\alpha $. 
In Fig. \ref{Chi2_to_zero} we compare the performance of the two values 
$\alpha = (1/2)\lambda_C=5/8$ (left panel) and $\alpha =1/2$ (right panel).
While in the former case one does observe data collapse (small preasymptotic corrections
are only observed for the smaller values of $s$, as expected) and linear behaviors,
both these feature are clearly lost in the latter case.
Notice also, as a further confirmation of the accuracy of the determination $\alpha = 5/8$,
that in the right panel the curves for small $C$ can be well fitted by the power law 
$\sim C^{5/4}$ (bold-dotted green line), as expected on the basis of Eq. (\ref{anticheck})
with $\alpha =5/8$ and $\beta = 1/2$.

As already stated, the value $\alpha =5/8$ is in contrast with the one $\alpha =1/2$
which was argued before.
However, previous computations were always carried out in quenches to finite
final temperatures (indeed, as we will see in the next sections, we recover 
a smaller value of $\alpha $ - compatible with the one ($\alpha=0.5-0.56$) found in the aforementioned
literature, when considering quenches to $T_f>0$), whereas to the best of our knowledge 
this is the first determination of this exponent in the case with $T_f=0$.
We will comment in Secs. \ref{to15}, \ref{to22} on this new result and we will provide 
a possible interpretation of the discrepancy between the present determination of $\alpha $ and
the previous ones.

\subsubsection{From $T_i=T_c$} \label{TcT0}

Let us now investigate which differences occur in the scaling properties of the system when the
quench is made from the critical point $T_i=T_c$ instead of having $T_i=\infty$.
The behavior of $L(t)$ in this case is shown in Fig. \ref{L_to_zero} (upper set of 
curves).
Here it is seen that $L$ is considerably larger than the one computed in the
quench from $T_i=\infty$. This can be understood since the system at infinite
temperature is maximally disordered while the critical state is a coherent one, although 
without order. 
Larger values of $L(t)$ and -- possibly --
the strong correlations present in the initial critical state, bring in finite-size
effects at earlier times as compared to the quench from $T_i=\infty$.
Indeed also in this case the expected power-law $L(t)\sim t^{1/z}$ with $z=2$ 
sets in around $t\simeq 10$ but, differently from the case with $T_i=\infty$,
one observes an upward bending of the curves starting from $t\simeq 10^3$ onwards.
The bending is more pronounced for smaller system sizes, confirming that it occurs earlier and
indicating that we are in the presence of important finite-size effects for 
$t\gtrsim 10^3$. 
Notice that finite-size effects do not produce in this case
an abrupt modification with respect to the behavior in an infinite system but,
rather, a gradual drift (in this case an upward raising) which can be confused with
a genuine effect. For instance, on the basis of Fig. \ref{L_to_zero} 
one could erroneously conclude that the asymptotic exponent is smaller than $z=2$.
Clearly, finite-size effects not only modify the behavior of $L(t)$ 
but affect all observables quantities, as we will discuss below. 

Starting from the equal-time correlation function, shown in 
Fig. \ref{G_to_zero} (upper group of curves), one finds 
data collapse of the curves at different times
by plotting $G$ against the rescaled space $y=r/L(t)$, as expected on the basis of Eq. (\ref{scalG}), 
but this is only true 
in a small-$r$ region which
shrinks as time increases. The breakdown of dynamical scaling at large $r$ is another 
clear manifestation of the finite-size effects. 
In an infinite system the curves of Fig. \ref{scalG} would collapse for any 
value of $r$ and at any time (except at such early times that scaling has not yet set in).
In our finite system this occur only in a range which gets narrower as $L(t)$ approaches 
$\Lambda$. Despite this, 
when time is sufficiently small there is room
to observe the typical large-distance power-law behavior $g(y)\sim y^{-\eta}$, with $\eta =1/4$, of 
Eq. (\ref{critg}) induced by the reminiscence of the initial critical state. 

The autocorrelation function is plotted against $x=L(t)/L(s)$
in Fig. \ref{C_to_zero}.
For all values of $x$ the collapse expected on the basis of Eq. (\ref{scalC}) is worse than
the one observed in the quench from $T_i=\infty$
(the effect is partly masked in Fig. \ref{C_to_zero} because data are compressed).
This is probably due to the combined effect of short-time corrections and finite-size
effects. Nevertheless the behavior $C(x)\sim x^{-\lambda_C}$ expected for large $x$ 
on the basis of Eq. (\ref{largec}) with
$\lambda _C=1/8$ is well reproduced in the regime of relatively large times but
such that finite-size effects are still negligible (namely, as already noticed
discussing the data for $L(t)$ (i.e. Fig. \ref{L_to_zero}), for $t\lesssim 10^3$
(for the curve with $s=10$ this roughly amounts to $x\lesssim 10$).
For larger values of $x$ the curves for $C(t,s)$ bend upwards, similarly to what
observed for $L(t)$. Needless to say, the behavior of $C$ is profoundly different
when the quench is made from $T_i=\infty$ or from $T_i=T_c$.

Next we consider the response function. To the best of our knowledge the computation
of this quantity 
in a quench from criticality was never performed before.
The behavior of $\chi$ in this case is shown in Fig. \ref{Chi_to_zero} (lower set of curves).
The curves can be collapsed, according to Eq. (\ref{scalChi}),
by plotting $L(s)^{\alpha}\chi (t,s)$ against $x=L(t)/L(s)$ where,
recalling the previous discussion, the exponent $\alpha $ is expected
to be equal to $\lambda _\chi$. The data of Fig. \ref{scalChi} show that also in this case 
$\alpha $ is well consistent with the
value $\alpha =5/8$, namely the same value obtained in the quench from infinite temperature.
This indicates that, while the behavior of the autocorrelation $C$ is profoundly different in
the two cases with $T_i=\infty$ and $T_i=T_c$, as to give $\lambda _C$ exponents as different
as $\lambda _C=5/4$ and $\lambda _C=1/8$, the response function exponent is insensitive
to the properties of the initial state. An interpretation of this fact will be given in 
Sec. \ref{largen} where the exact solution of the large-$N$ model, which shares the same property, will be
discussed.

In the following we will show how thermal fluctuations occurring when $T_f\neq 0$ modify
the scaling picture found above for the zero-temperature quench.  We will consider the two cases
with an intermediate temperature $T_f=1.5\simeq 0.66 T_c$ and one, $T_f=2.2\simeq 0.97 T_c$, close to
the critical one. 

\subsection{Quenches to $T_f=1.5$} \label{to15}

\subsubsection{From $T_i=\infty$} \label{TinfT15}

\begin{figure}
\vspace{1cm}
\rotatebox{0}{\resizebox{.85\textwidth}{!}{\includegraphics{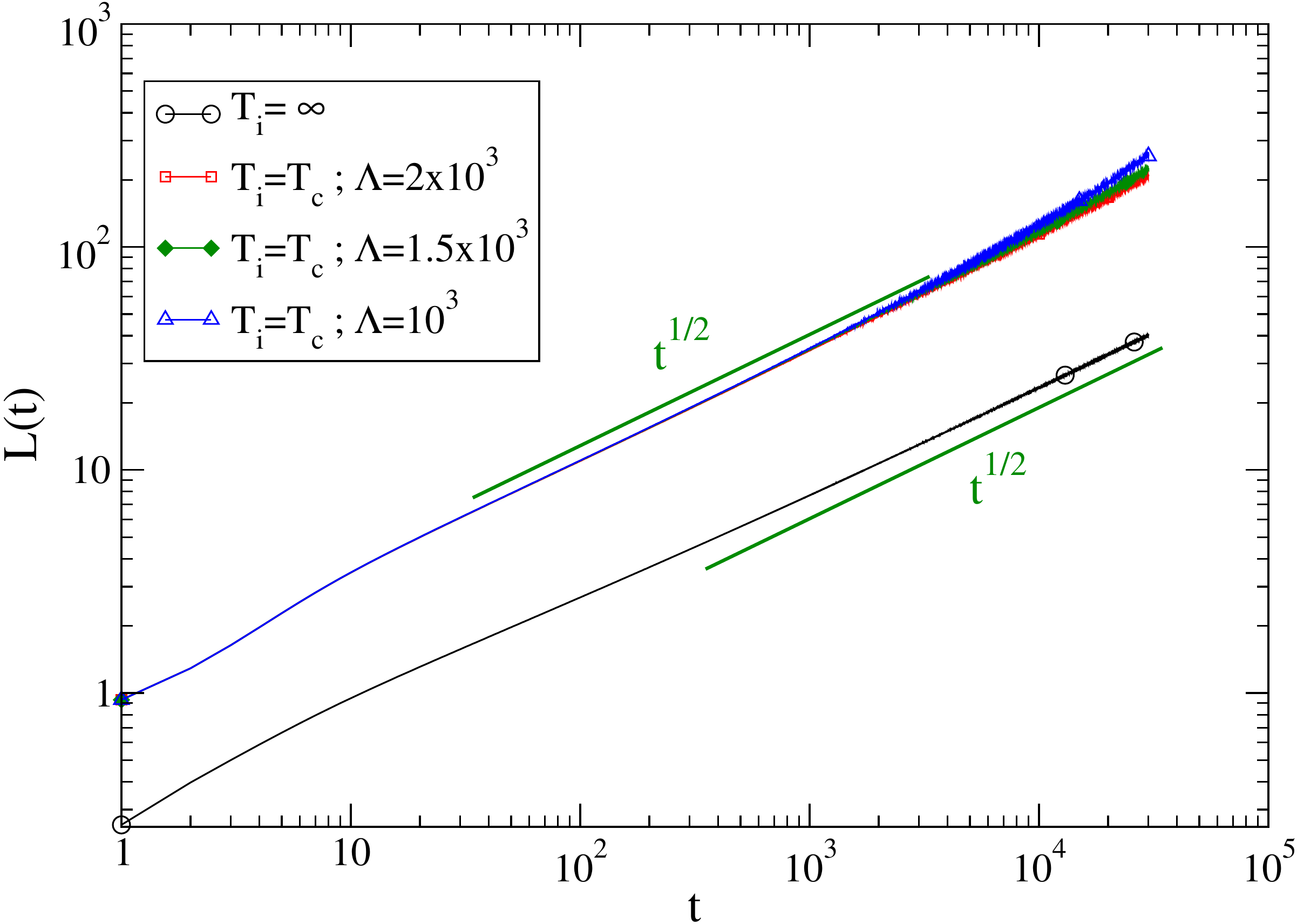}}}
\caption{(Color on-line). $L(t)$ is plotted against $t$ on a log-log plot
for the quench of
the system to $T_f=1.5$ starting from the equilibrium state at $T_i=\infty$
(lower black curve with a circle)
and at $T_i=T_c$ (upper partially collapsing three curves, corresponding to three different
system sizes $\Lambda $ as detailed in the key). The bold green lines are the behavior
$L(t)\sim t^{1/2}$ expected for an infinite system for large times.}
\label{L_to_15}
\end{figure}

As shown in Fig. \ref{L_to_15} the behavior of $L(t)$ in the quench to $T_f=1.5$ 
is very similar to the case with $T_f=0$ discussed previously. 
The expected power-law $L(t)\sim t^{1/z}$
with $z=2$ (a fit in the last decade provides $z=2.031$) sets in quite early 
and there is no indication of any finite-size effect. 

\begin{figure}
\rotatebox{0}{\resizebox{.85\textwidth}{!}{\includegraphics{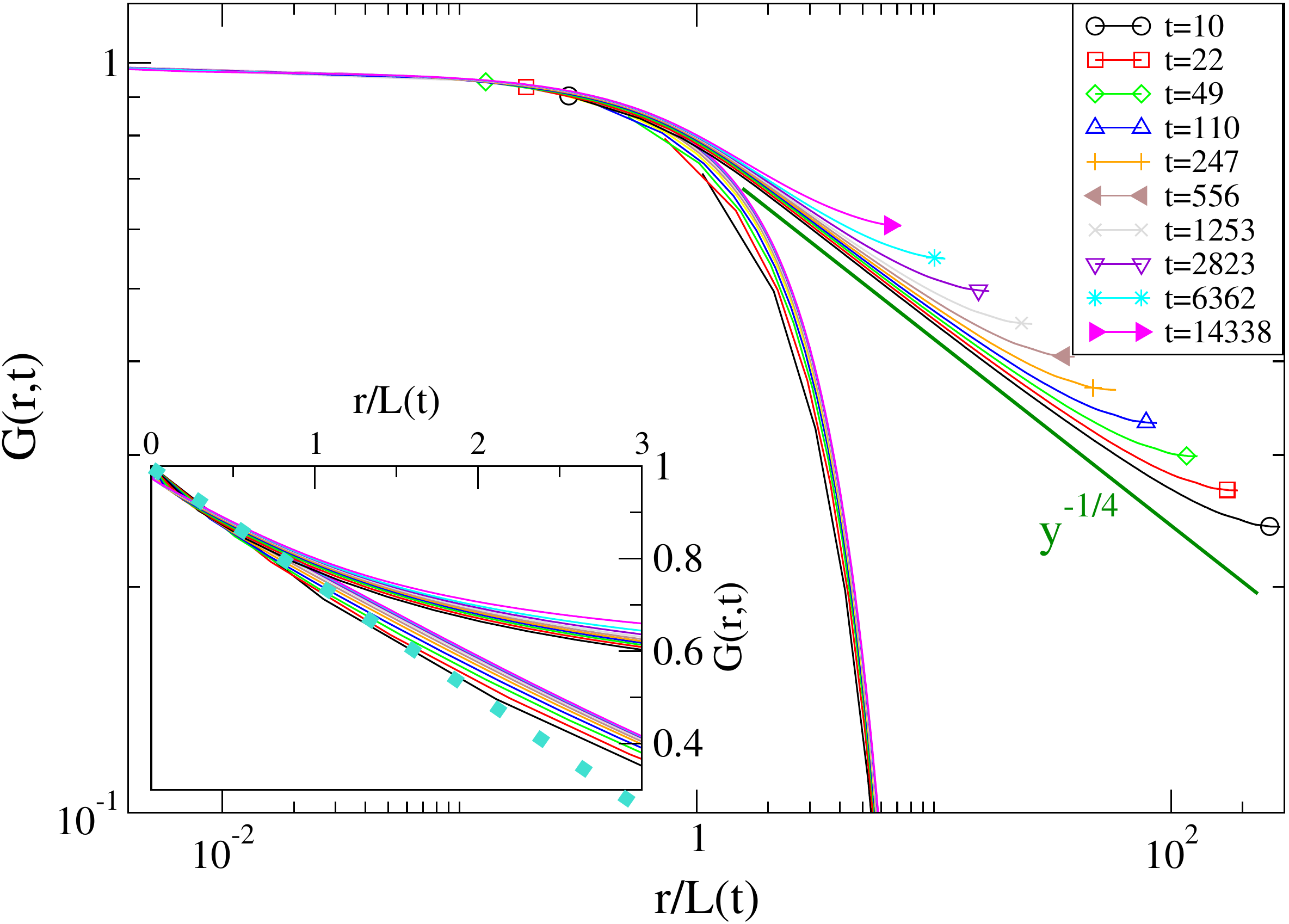}}}
\caption{(Color on-line). The correlation function $G(r,t)$ for a quench to $T_f=1.5$ 
is plotted against $y=r/L(t)$ 
on a double-logarithmic plot at the different times indicated in the key
(these are determined as to be exponentially spaced by an automatic routine ). 
The lower set of collapsing curves (without symbols), correspond to the case 
discussed in Sec. \ref{TinfT15} of a quench from an equilibrium state at the initial
temperature  $T_i=\infty$. The upper group of curves,
marked with a symbol, correspond to the case discussed in Sec. \ref{TcT15} of  a quench
starting from the critical state, i.e. $T_i=T_c$.
The bold-green straight line is the power-law $y^{-1/4}$ of Eq. (\ref{critg}).
In the inset a zoom of the same data in the region of small $r/L(t)$ is plotted on a 
double-linear plot. The bold-dotted turquoise line is the linear behavior 
of Eq. (\ref{porod}), namely the Porod's law.}
\label{G_to_15}
\end{figure}

The behavior of the equal-time correlation function is shown in 
Fig. \ref{G_to_15} (lower group of curves) for various times (see key)
on a double-logarithmic plot.
As expected on the basis of Eq. (\ref{scalG}), data collapse of the
curves at different times is obtained 
by plotting $G$ against the rescaled space $y=r/L(t)$ in all the region where 
$G$ is significantly larger than zero.  The collapse is poor at small times but it gets
better moving to larger $t$.
As it can be seen by zooming the small-$r$ region
in the inset of Fig. \ref{G_to_15} with linear axes,
the Porod's law (\ref{porod}) is well reproduced also in this quench.

\begin{figure}
\rotatebox{0}{\resizebox{.85\textwidth}{!}{\includegraphics{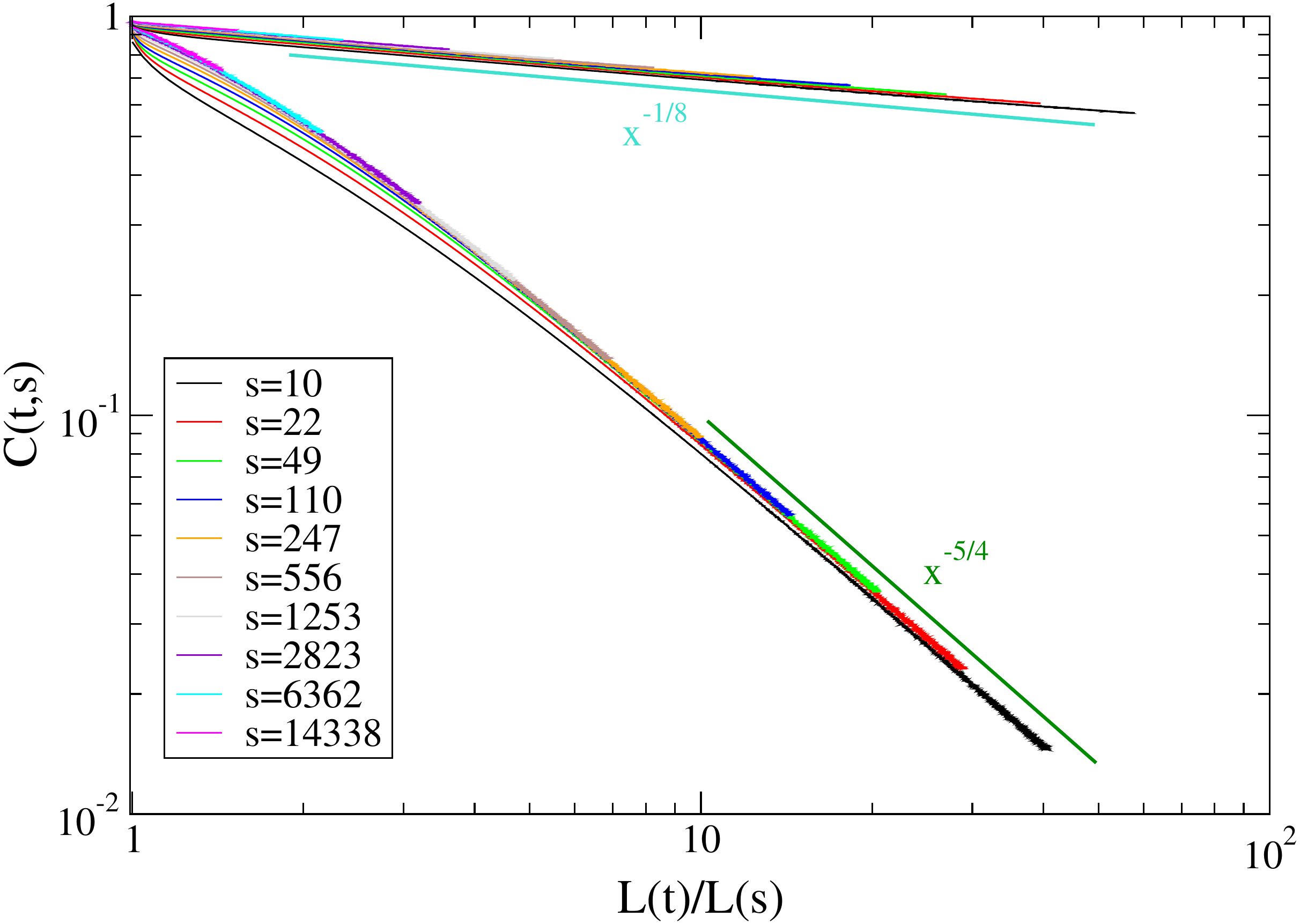}}}
\caption{(Color on-line). $C(t,s)$ is plotted against $x=L(t)/L(s)$ 
for the quench of
the system to $T_f=1.5$ starting from the equilibrium state at $T_i=\infty$
(lower set of curves)
and at $T_i=T_c$ (upper group of curves) for different times (see key).
The bold dark-green and turquoise lines are the expected 
power-laws $x^{-\lambda _C}$ with $\lambda _C=5/4$ and $\lambda _C=1/8$
for the quenches from $T_i=\infty$ and from $T_i=T_c$, respectively.}
\label{C_to_15}
\end{figure}

The autocorrelation function is plotted against $x=L(t)/L(s)$
in Fig. \ref{C_to_15}. Also for this quantity
the collapse expected on the basis of Eq. (\ref{scalC}) is poor for the smaller values
of $s$ but becomes progressively more accurate, particularly in the region of large $x$, as $s$ is increased. 
The expected behavior $C(x)\sim x^{-\lambda_C}$ of 
Eq. (\ref{largec}) with
$\lambda _C=5/4$ is quite well
observed (a fit of the curve with $s=10$ for $x\ge 30$ gives $\lambda _C=1.236$).

\begin{figure}
\rotatebox{0}{\resizebox{.85\textwidth}{!}{\includegraphics{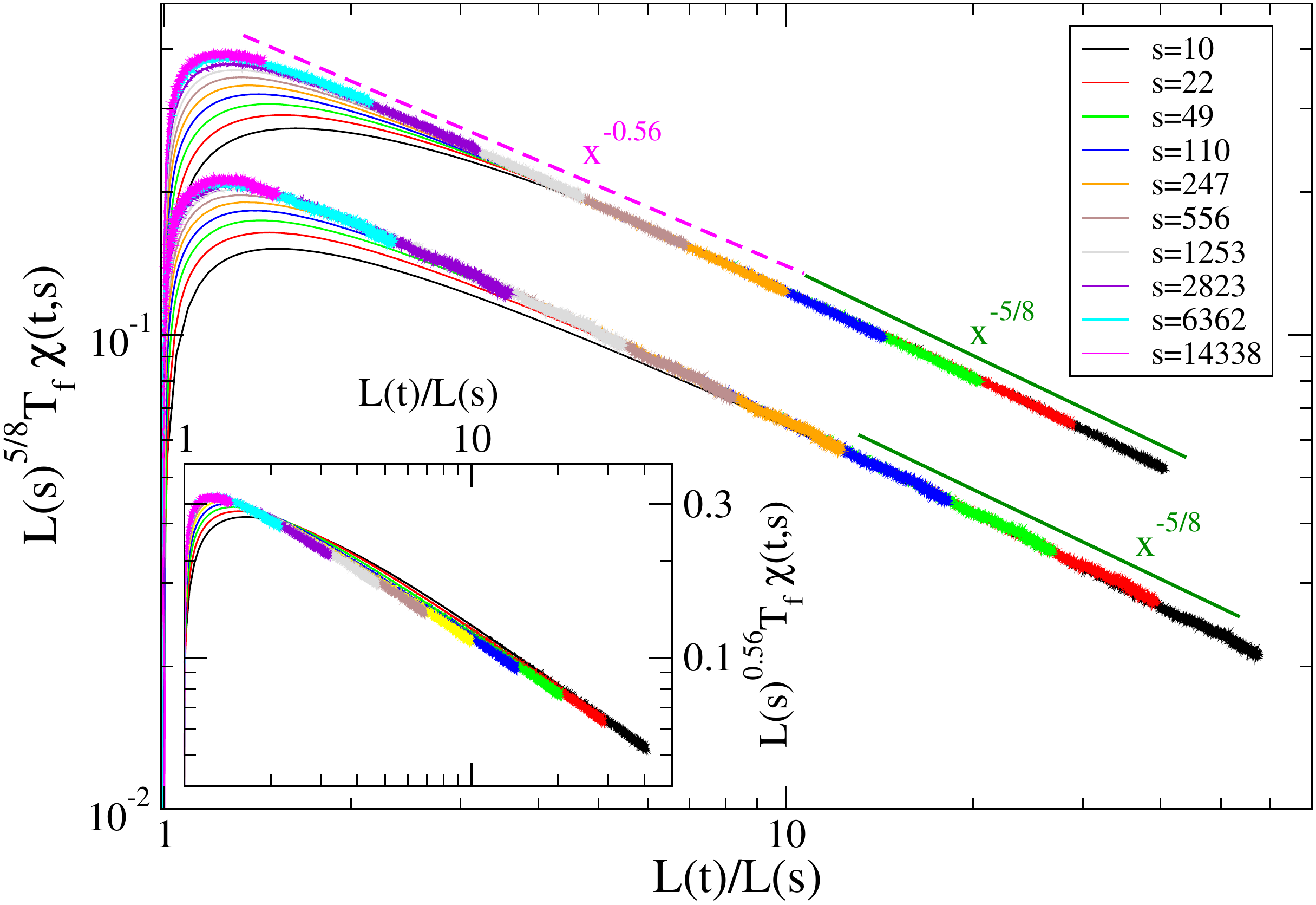}}}
\caption{(Color on-line). $L(s)^{\frac{5}{8}}T_f\,\chi(t,s)$ is plotted against $x=L(t)/L(s)$ 
for the quench of
the system to $T_f=1.5$ starting from the equilibrium state at $T_i=\infty$
(upper set of curves)
and at $T_i=T_c$ (lower group of curves) for different times (see key).
The bold dark-green and dashed-magenta lines are the
power-laws $x^{-\lambda _\chi}$ with $\lambda _\chi=5/8$ and $\lambda _\chi=0.56$.
In the inset, starting from the same data of the main figure for $T_i=\infty$, we plot
$L(s)^{0.56}T_f \chi(t,s)$ against $L(t)/L(s)$.}
\label{Chi_to_15}
\end{figure}

The response function is shown in Fig. \ref{Chi_to_15}.
This quantity was previously computed several times \cite{unquarto} for a quench to the same 
final temperature $T_f=1.5$ considered here. The present study therefore
allows us to compare
our results with previous ones and to discuss the discrepancy on the $\alpha $ exponent found 
in the previous section \ref{to0}.
According to Eq. (\ref{scalChi}) one should get 
the collapse of curves for $\chi $ at different waiting times $s$ 
by plotting $L(s)^{\alpha}\chi (t,s)$ against $x=L(t)/L(s)$,
where the exponent $\alpha$ equals the exponent $\lambda _\chi$
defined in Eq. (\ref{largeh}).
Comparison of the large-$x$ behavior of the curves in Fig. \ref{Chi_to_15}
with the dark-green line $x^{-5/8}$ shows that $\lambda_\chi$ is 
very well compatible with the value $\lambda _\chi=5/8$ found with $T_f=0$
also in this finite-temperature quench. 

In previous studies \cite{unquarto,revresp,henkelrough} the scaling of the response function was
usually written in terms of the time variable as
\be
\chi(t,s)=s^{-a}\widetilde h\left [\frac{t}{s}\right ].
\label{scalChit}
\ee
Assuming the asymptotic behavior $L(t)\propto t^{\frac{1}{2}}$ this implies
$a =\frac{\alpha}{2}$. In \cite{unquarto} it was found
$a\simeq 0.28$ which implies $\alpha \simeq 0.56$.  
Recalling that $\alpha=\lambda _\chi$, 
in Fig. \ref{Chi_to_15} it is observed that the exponent $\lambda _\chi =0.56$ 
describes reasonably well the $x$-dependence of the scaling function $h$ in 
Eq. (\ref{scalChi}) up to intermediate values of $x=L(t)/L(s)\simeq 10$.
We will comment later on the occurrence of such intermediate behavior.
For larger values the curves bend slightly and progressively downward
and the exponent $\lambda _\chi=5/8$ looks more consistent with the data
(a fit in the region with $x\ge 10$ provides $\lambda _\chi=0.63$).  
Notice that the intermediate exponent $\alpha \simeq 0.56$ was not observed
in the quench to $T_f=0$, signaling that this is due to thermal fluctuations.

\begin{figure}
\rotatebox{0}{\resizebox{.45\textwidth}{!}{\includegraphics{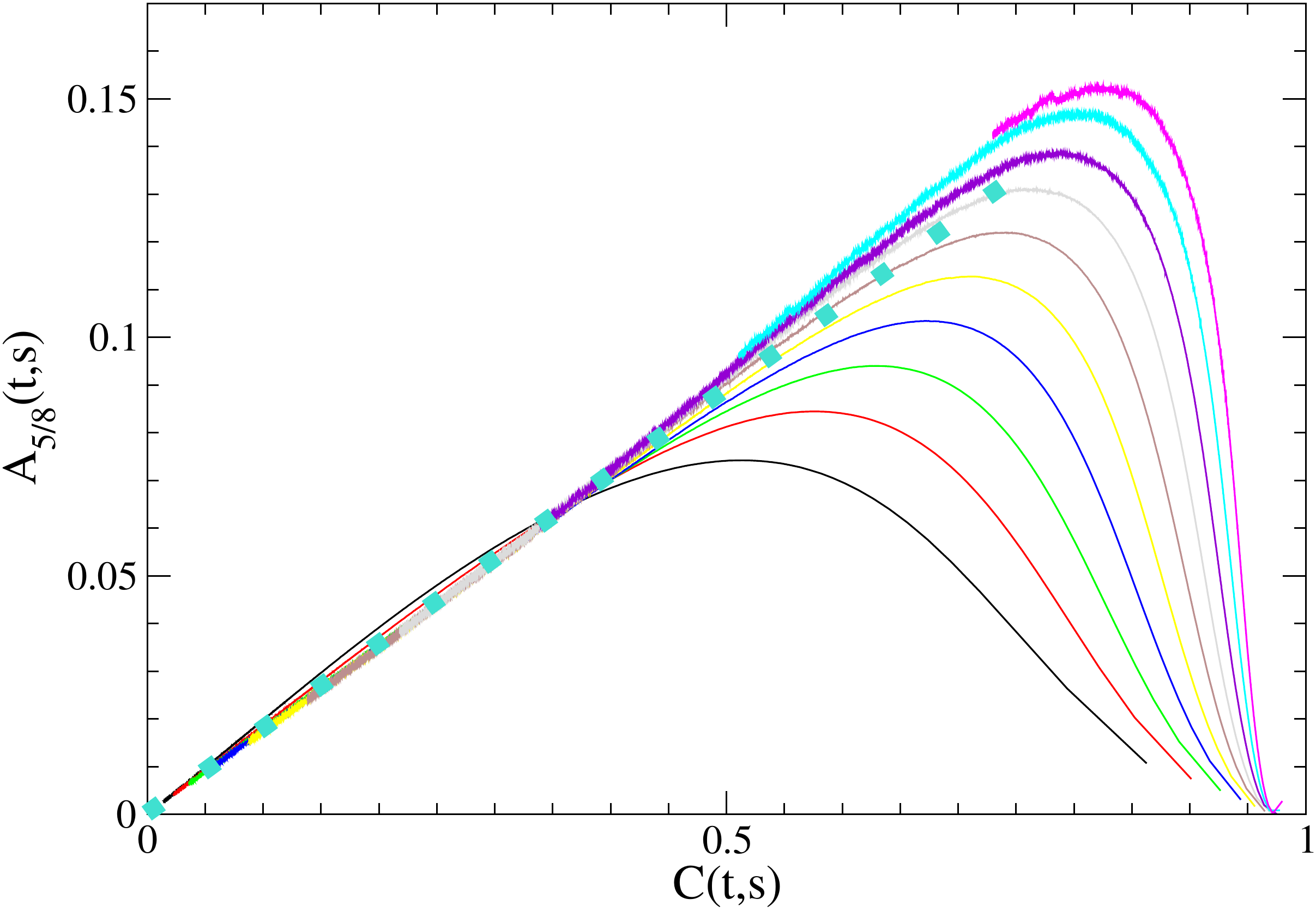}}}
\rotatebox{0}{\resizebox{.45\textwidth}{!}{\includegraphics{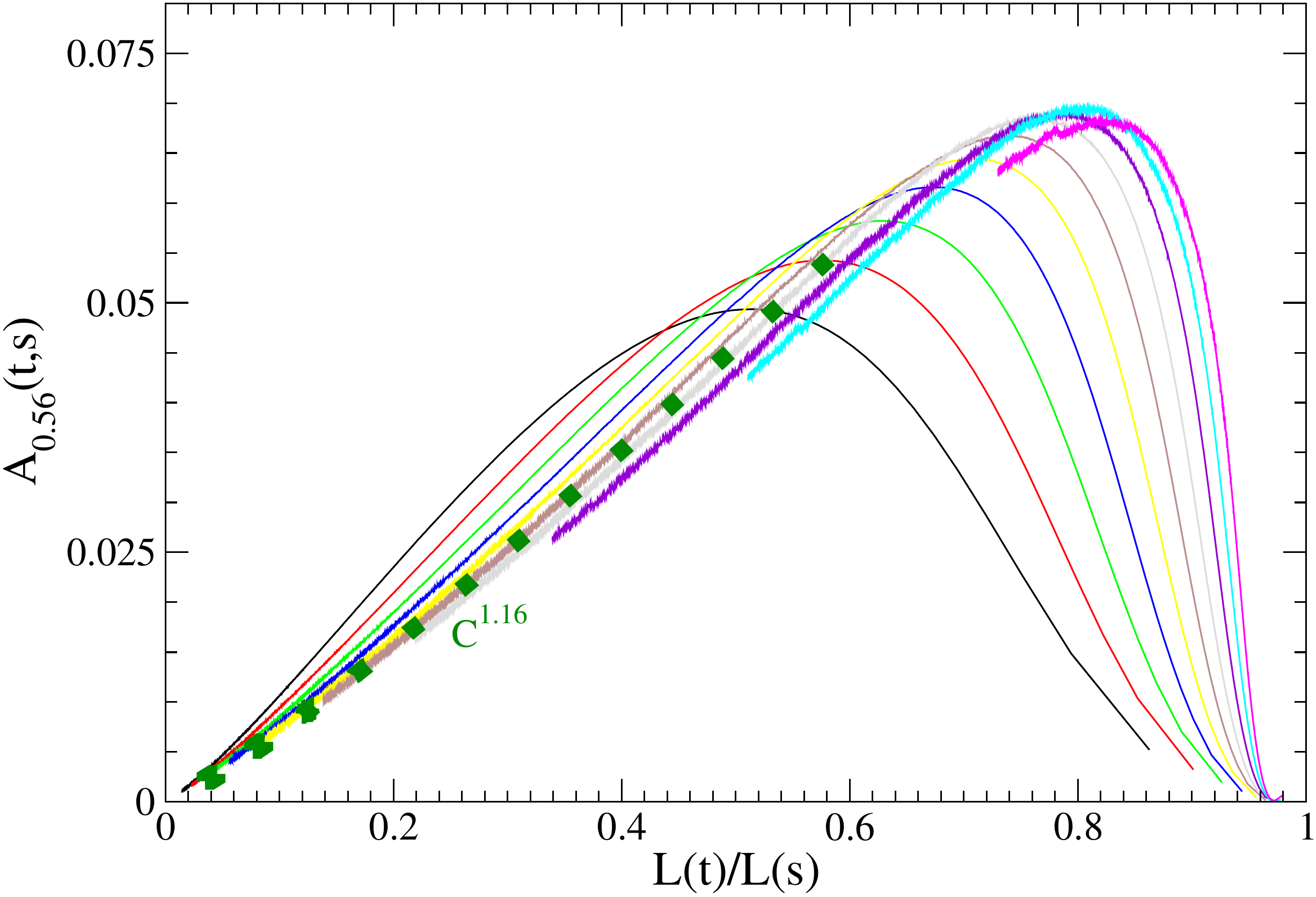}}}
\caption{(Color on-line). The function $A_{\frac{5}{8}}(t,s)$ (left panel)
and $A_{0.56}(t,s)$ (right panel)
are plotted against $C(t,s)$ for the quench of
the system to $T_f=1.5$ starting from the equilibrium state at $T_i=\infty$
for different times (see key).
The bold-dotted turquoise line on the left panel is guideline for the linear behavior
$[L(s)^{5/8}\chi (t,s)]^2\propto C(t,s)$. The bold-dotted green one on the right panel 
is the behavior $C(t,s)^{1.16}$.}
\label{Chi2_to_15}
\end{figure}

The behavior of the response function suggests that the exponent 
$\alpha \simeq 0.56$ could be due to a 
preasymptotic mechanism associated to thermal fluctuations, while the truly asymptotic behavior
is the one with $\alpha=\lambda_\chi=(1/2)\lambda _C$ as in the quench to $T_f=0$.
Indeed, when trying to obtain data collapse by plotting $L(s)^\alpha \chi(t,s)$
against $L(t)/L(s)$ one finds that curves superimpose better with $\alpha =5/8$
(main figure) than with $\alpha = 0.56$ (inset), although a satisfactory collapse is obtained
in both cases.
This is confirmed in Fig. \ref{Chi2_to_15} where we compare the performance of the two values 
$\alpha = 5/8$ (left panel) and $\alpha =0.56$ (right panel) with the method of Sec. \ref{TinfT0},
by using Eq. (\ref{check}).
With $\alpha=5/8$ one observe both data collapse (with small preasymptotic corrections
for small $s$, similarly to the case with $T_f=0$) and linear behavior 
$A_{\frac{5}{8}}\sim C$ in the small-$C$ region. On the other hand
with $\alpha=0.56$ both these feature are clearly lost. 
Furthermore, in the right panel the curves for small $C$ can be well fitted 
by the power law $\sim C^{1.16}$ (bold-dotted green line), as expected on the basis of Eq. (\ref{anticheck})
with $\alpha =5/8$ and $\beta = 0.56$.

Let us now comment on a possible interpretation of the behavior of $\chi $ and,
in particular, of the $\alpha $-exponent. As already mentioned the measured value $\alpha=0.5-0.56$
was interpreted as a result of a putative exponent $\alpha=1/2$ associated to the kinetic roughening of the
domains interfaces whose width is expected to scale as 
\be
\lambda (t) \simeq A(T_f)L(t)^{\alpha _R},
\label{rough}
\ee
where $\alpha _R$ is the so-called {\it roughness exponent} ($\alpha _R=1/2$ in $d=2$) and 
$A(T_f)$ is an increasing function of temperature. It is known that 
the extra length $\lambda $ can produce pre-asymptotic corrections to scaling in many observables,
as shown in \cite{interf2d}. However, for large times these corrections can be neglected
because $\lambda (t)$ is eventually dominated by $L(t)$. This is true, for instance, in quantities such as 
$G$ or $C$. On the other hand, the statement that the asymptotic
exponent $\alpha $ is the one ($\alpha =1/2$) associated to roughness amounts to assume that 
the mechanism whereby the response is built relies only on $\lambda $,
despite the fact that $L\gg \lambda$ is the dominant length. It is interesting to notice that, since
roughness is expected to vanish at zero temperature, i.e. $A(T_f=0)=0$, this mechanism
cannot be sustained in a zero-temperature quench. Indeed we have shown that in this case a different
exponent $\alpha =\lambda _C/2=5/8$ is very neatly observed. The study of the quench to $T_f=1.5$
presented here, upon extending the range of simulated times with respect to the previous ones, 
allows us to argue that the value $\alpha =5/8$ is the correct asymptotic one also in a finite-temperature
quench, while a smaller value $\alpha \simeq 0.56$ is only observed pre-asymptotically. 
Early-time corrections to the response function are a well known fact and are discussed in \cite{resp2d}. 
Notice that
the crossover from the pre-asymptotic roughness-related mechanism to the truly asymptotic one is 
regulated by $A(T_f)$: the larger is $T_f$, the larger is $A$ and this makes the pre-asymptotic
behavior with $\alpha =1/2$ to last longer. Finally, let us comment on the fact that the measured 
exponent $\alpha$ has been always found larger that $\alpha =1/2$ signaling that also in the
previous determinations the crossover towards $\alpha =5/8$ was very probably already present. 

\vspace{2cm}

\subsubsection{From $T_i=T_c$} \label{TcT15}

In the case of a quench to the finite final temperature $T_f=1.5$ the differences 
between an infinite initial temperature $T_i=\infty$ and a critical one $T_i=T_c$
occur similarly to what observed in the quench to $T_f=0$.
In particular $L(t)$, see Fig. \ref{L_to_15} (upper set of 
curves), is considerably larger than the one for $T_i=\infty$
and finite-size effects start occurring around $t\simeq 10^3$ with the
modalities discussed in Sec. \ref{TcT0}.
The behavior of $G(r,t)$, shown in Fig. \ref{G_to_15} (upper group of curves), is also very
similar to the one discussed in Sec. \ref{TcT0}, with scaling obeyed (according to Eq. (\ref{scalG}))
for sufficiently small values of $y=r/L(t)$, a remnant of the critical behavior
of Eq. (\ref{critg}) for an intermediate range of $y$ and marked finite-size effects at large-$y$.

Also the autocorrelation function, plotted against $x=L(t)/L(s)$ 
in Fig. \ref{C_to_zero} (upper set of curves), closely follows the behavior 
already observed in Sec. \ref{TcT0} for a quench to $T_f=0$, with the difference 
that the quality of the scaling collapse is worse than before.
Despite this, the behavior $C(x)\sim x^{-\lambda_C}$ expected for large $x$ 
on the basis of Eq. (\ref{largec}) with
$\lambda _C=1/8$ is well reproduced in an intermediate regime of $x=L(t)/L(s)$ where
finite-size effects are not important.

Analogous considerations apply to the response function, which is shown in Fig. \ref{Chi_to_15}
(lower group of curves). This quantity behaves very similarly to the quench from $T_i=\infty$ to
$T_f=1.5$ discussed in Sec. \ref{TinfT15}. In particular one has a good indication of an exponent
$a=\lambda _\chi=5/8$, while a somewhat smaller value $\lambda _\chi \simeq 0.56$ is compatible with the data at earlier times.

In conclusion, for all the quantities considered we do not find significant differences between 
a quench to $T_f=0$ or to $T_f=1.5$ (starting both from $T_i=\infty$ and $T_i=T_c$),
apart from the quality of the scaling which gets worse upon raising $T_f$;
this confirms that $T_f$ is an {\it irrelevant parameter} \cite{brayren} in a renormalization group sense.
Concerning the role of the initial temperature, $\lambda _C$ turns out to be markedly 
influenced by $T_i=\infty$ ($\lambda _C=5/4$) or $T_i=T_c$ ($\lambda _C=1/8$), 
both with $T_f=0$ and $T_f=1.5$. On the contrary, the response function exponents 
are basically independent on $T_i$. Recalling that the memory of the initial condition is 
retained at large wavelength this suggests that the contribution of large scales to the response
function is negligible whereas it is important for the autocorrelation.

\subsection{Quenches to $T_f=2.2$} \label{to22}

\subsubsection{From $T_i=\infty$} \label{TinfT22}

\begin{figure}[h]
\vspace{1cm}
\rotatebox{0}{\resizebox{.85\textwidth}{!}{\includegraphics{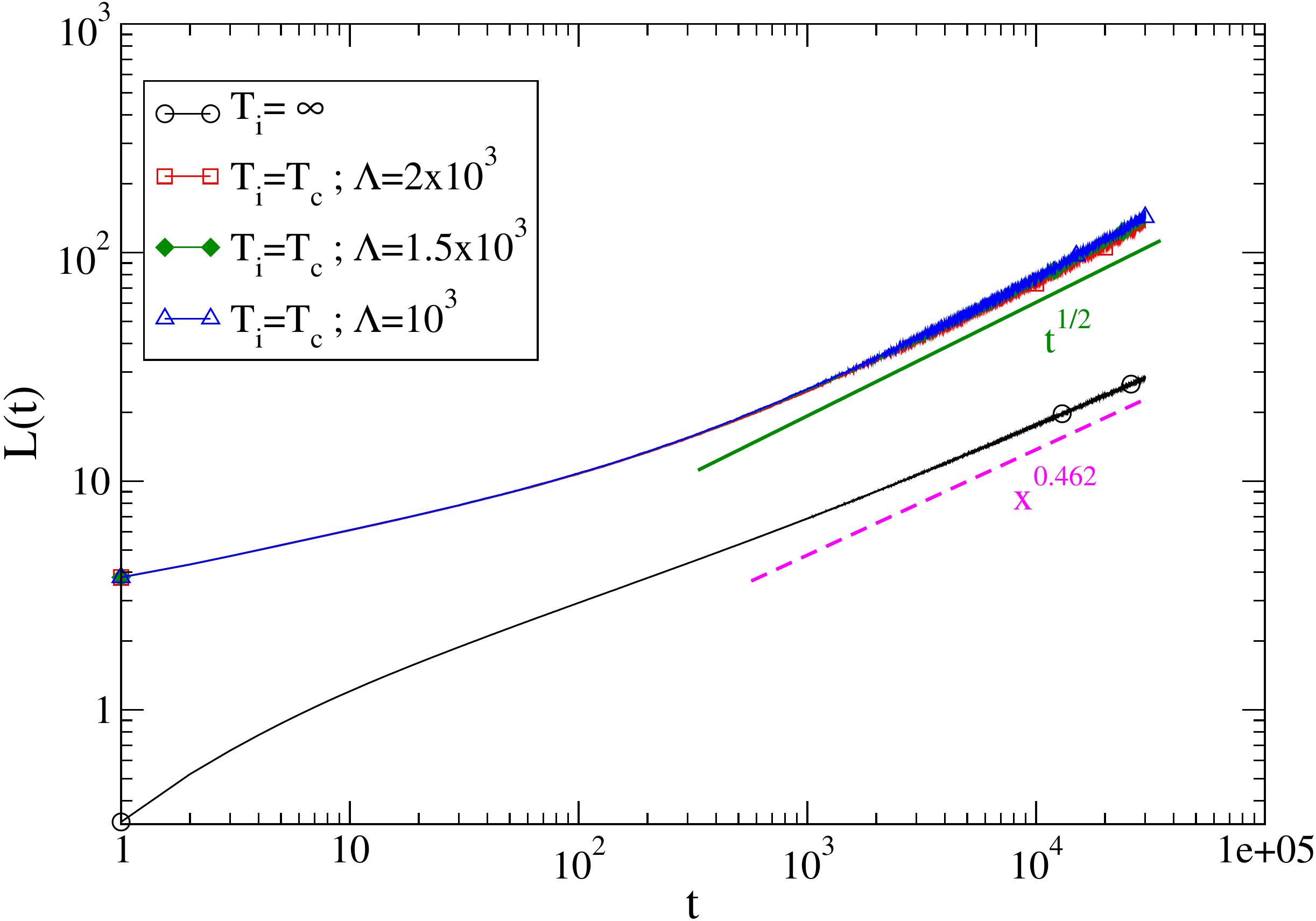}}}
\caption{(Color on-line). $L(t)$ is plotted against $t$ on a log-log plot
for the quench of
the system to $T_f=2.2$ starting from the equilibrium state at $T_i=\infty$
(lower black curve with a circle)
and at $T_i=T_c$ (upper partially collapsing three curves, corresponding to three different
system sizes $\Lambda $ as detailed in the key). The bold green lines are the behavior
$L(t)\sim t^{1/2}$ expected for an infinite system for large times. The dashed-magenta line
is the behavior $t^{}$ expected in a critical quench.}
\label{L_to_22}
\end{figure}

\begin{figure}
\rotatebox{0}{\resizebox{.75\textwidth}{!}{\includegraphics{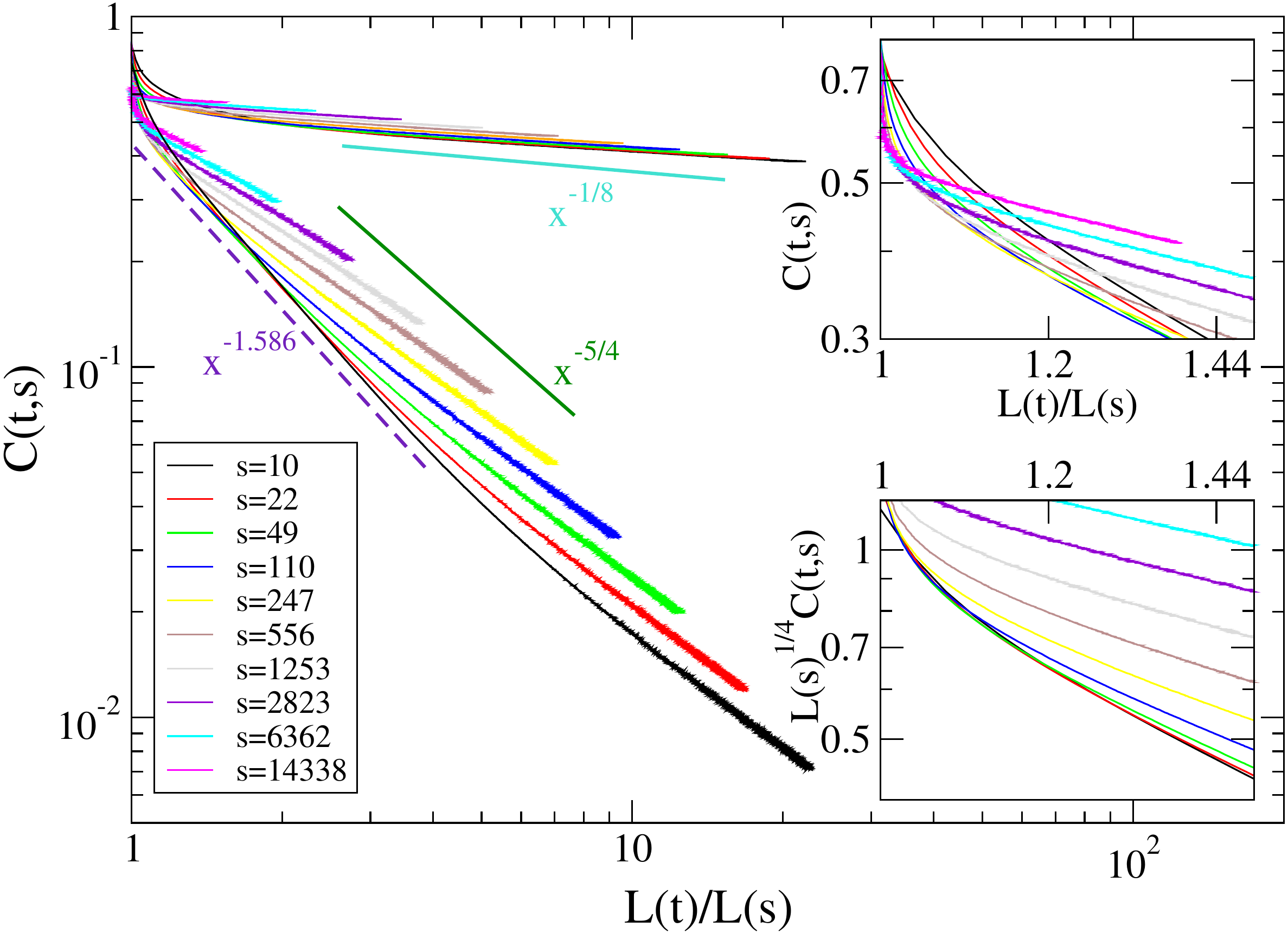}}}
\caption{(Color on-line). $C(t,s)$ is plotted against $x=L(t)/L(s)$ 
for the quench of
the system to $T_f=2.2$ starting from the equilibrium state at $T_i=\infty$
(lower group of curves)
and at $T_i=T_c$ (upper set of curves) for different times (see key).
The bold dark-green and turquoise lines are the expected 
power-laws $x^{-\lambda _C}$ with $\lambda _C=5/4$ and $\lambda _C=1/8$
for the quenches from $T_i=\infty$ and from $T_i=T_c$, respectively.
The dashed-indigo line is the power-law $x^{-1.586}$ expected for a quench
from $T_i=\infty$ to the critical point $T_f=T_c$.
In the upper inset a zoom on the small-$x$ part of the same curves of the main figure 
is shown. The lower inset contains the same data of the upper one but the rescaled quantity
$L(s)^{1/4}C(t,s)$ is plotted.}
\label{C_to_22}
\end{figure}

When quenching to a temperature as near to the critical one $T_c\simeq 2.26$ as $T_f=2.2$,
preasymptotic effects are so strong to prevent the observation of the expected asymptotic scaling.
This can be seen already from the behavior of the typical length $L(t)$ which is shown in
Fig. \ref{L_to_22}. Here one sees that the growth is slower than the expected one and a fit
for $t\ge 10^4$ yields an effective exponent $1/z_{eff}\simeq 0.44$. Notice that this behavior is
not too far from to the one $L(t)\sim t^{1/z_c}$, with $z_c= 2.1667(5)$ (i.e. $1/z_c\simeq 0.46$),
expected in a critical quench, namely one from $T_i=\infty$ to $T_f=T_c$.
This suggests that the proximity of $T_f$ to the critical point might affect the
behavior of the system at early times. We will confirm that this is the case by studying the
behavior of the autocorrelation function, which is plotted against $x=L(t)/L(s)$
in Fig. \ref{C_to_22} (lower set of curves).
The collapse expected on the basis of Eq. (\ref{scalC}) is 
not observed in the range of times accessed in our simulations,
although a tendency of the curves to get closer is observed as time $s$ grows large.
The expected behavior $C(x)\sim x^{-\lambda_C}$ of 
Eq. (\ref{largec}) with
$\lambda _C=5/4$ is neither observed. In place of this one has a power-law behavior
$C\sim x^{-1.586}$ for the smaller values of $s$ in an intermediate region of $x$.
This is the expected \cite{noigamba} behavior in a critical
quench, namely the one from $T_i=\infty$ to $T_f=T_c$, for which
\be
C= L(s)^{-(d-2+\eta)}F\left [\frac{L(t)}{L(s)}\right ]
\label{critscal}
\ee
with the large-$x$ behavior 
\be
F(x)\sim x^{(\theta -1)z_c+2-d-\eta}.
\label{cmpl}
\ee 
Here $\theta = 0.383(3)$
is the so called {\it initial slip} exponent, $\eta=1/4$, and $L(t)\sim t^{1/z_c}$
(so that the numerical value of the exponent in Eq. (\ref{cmpl}) is $-1.586$). 
This suggests quite clearly that, when quenching to a 
temperature near to the critical one, data collapse is
delayed because of the influence of the critical point which attracts -- in a 
renormalization-group language -- the trajectory of the flow at early times.
Since however $T_f<T_c$, one should observe the same asymptotic
behavior as in a quench to $T_f=0$ if sufficiently large times could be reached.
Indeed, in Fig. \ref{C_to_22} one can notice that the critical behavior $x^{-1.586}$ is
lost for the larger values of $s$ and the curves tend to bend downward at large-$x$,
presumably approaching the expected asymptotic behavior $x^{-5/4}$ at times much
larger than those accessed in our simulations.
A further confirmation that the critical scaling (\ref{critscal}) is obeyed at short
times is given by the analysis of the small-$x$ behavior, which is shown in the two insets
of Fig. \ref{C_to_22}. The upper inset is only a zoom of the main figure and shows that
data collapse is never observed, for any choice of $s$. On the other hand, in the lower inset one sees that,
by plotting $L(s)^{\frac{1}{4}}C$ against $x$,
data collapse is obtained for the smallest values of $s$ in a region of
$x$ which shrinks by increasing $s$, 
as it is expected on the basis of the critical scaling (\ref{critscal}).

A behavior similar to the one discussed insofar for the autocorrelation function is displayed
by the equal time correlation function and
by the response function (not shown). Also for these quantities the data collapse expected on the basis of
the scalings (\ref{scalG},\ref{scalChi}) are not observed in the range of times accessed in our simulations
due to important pre-asymptotic effects.

\subsubsection{From $T_i=T_c$} \label{TcT22}

\begin{figure}[h]
\vspace{1cm}
\rotatebox{0}{\resizebox{.45\textwidth}{!}{\includegraphics{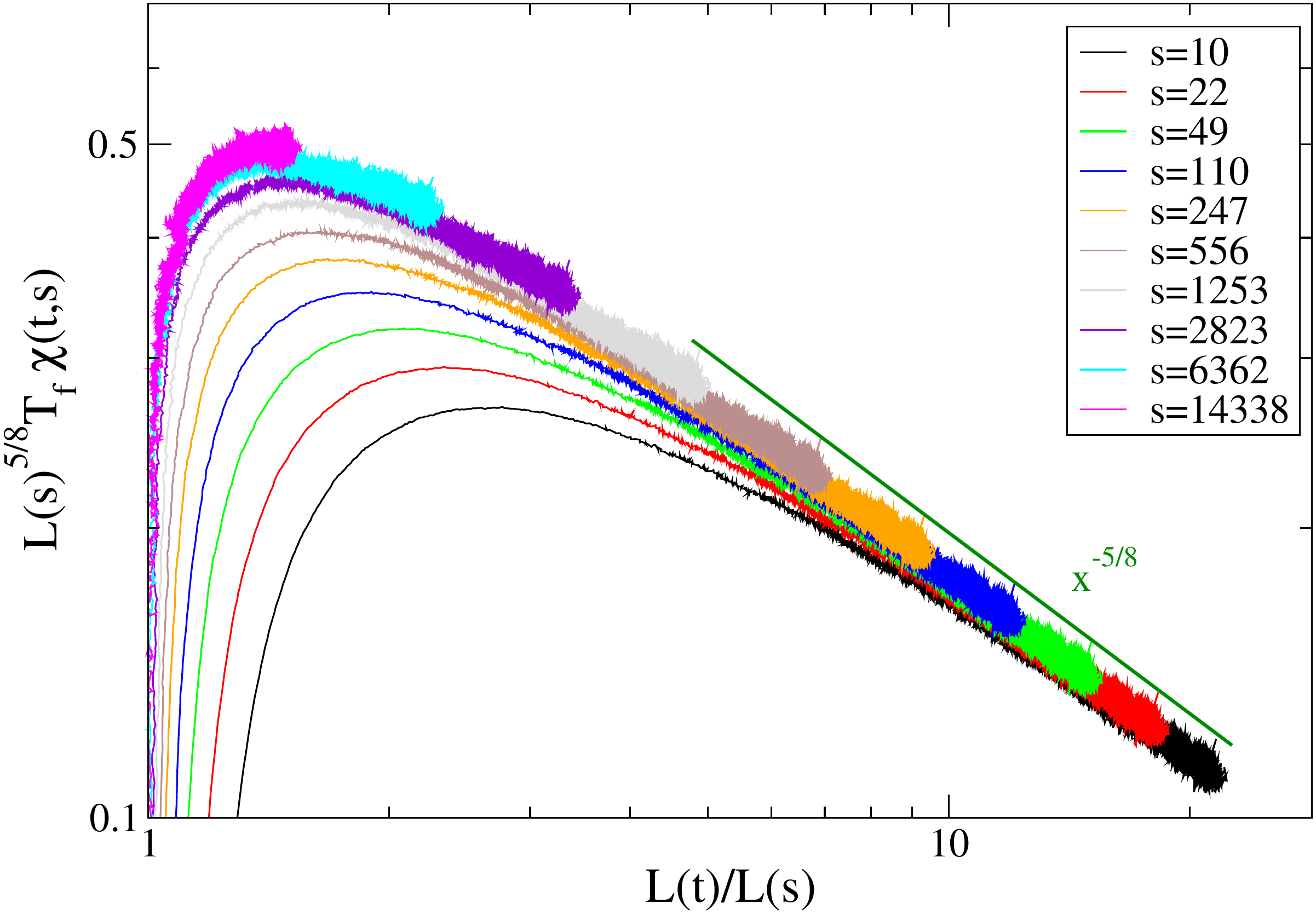}}}
\rotatebox{0}{\resizebox{.45\textwidth}{!}{\includegraphics{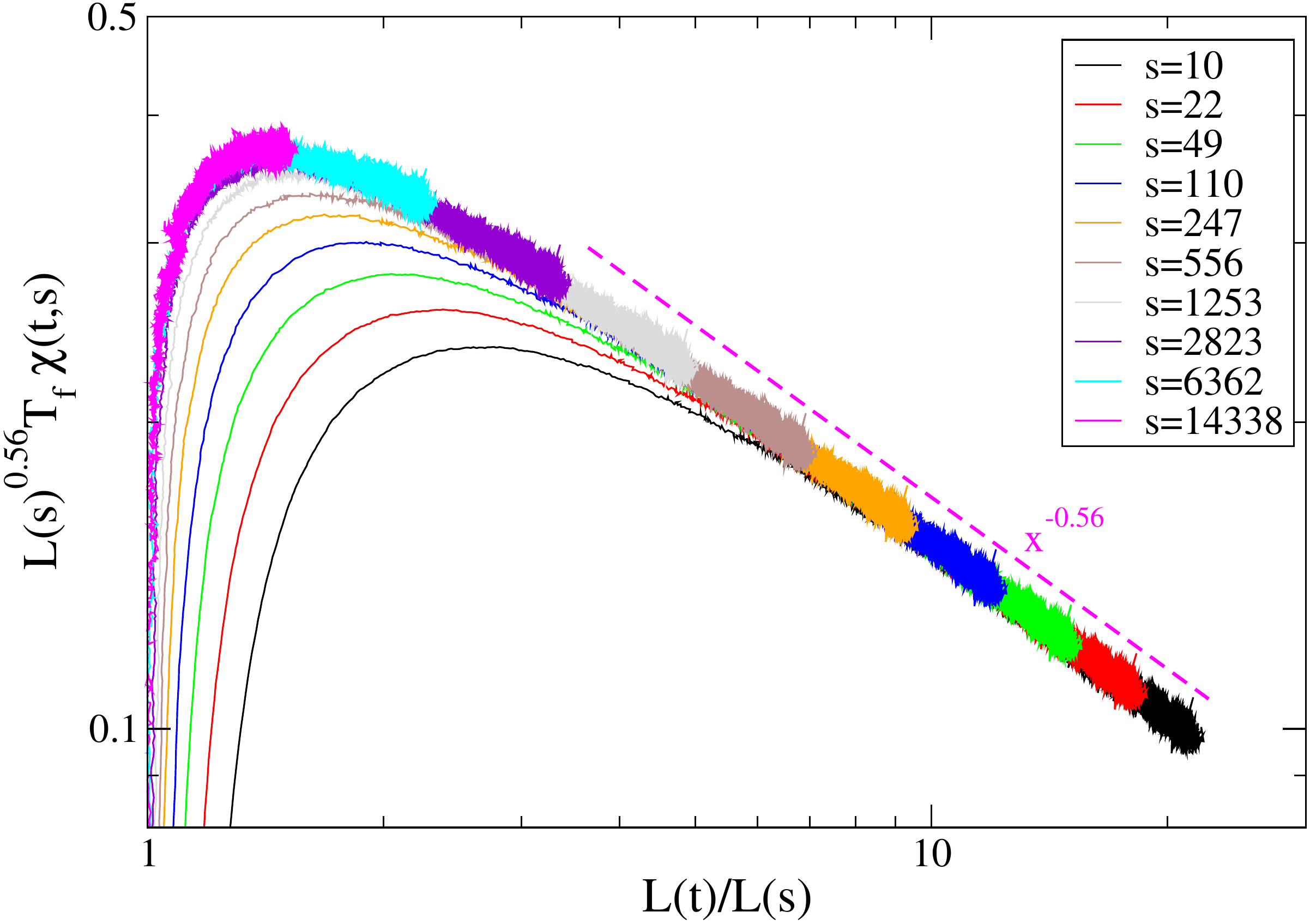}}}
\caption{(Color on-line). $L(s)^{\frac{5}{8}}T_f\,\chi(t,s)$ (left panel) or $L(s)^{0.56}T_f\,\chi(t,s)$
is plotted against $x=L(t)/L(s)$  for the quench of
the system to $T_f=2.2$ starting from the equilibrium state at $T_i=T_c$
for different times (see key).
The bold dark-green line (left panel) is the
power-law $x^{-\lambda _\chi}$ with $\lambda _\chi=5/8$.
The dashed magenta line (right panel) is the
power-law $x^{-\lambda _\chi}$ with $\lambda _\chi=0.56$.}
\label{Chi_to_22}
\end{figure}

The behavior of the characteristic domains size $L(t)$ is shown in Fig. \ref{L_to_22}.
Also in this case long-lasting pre-asymptotic corrections delay the asymptotic
behavior. Indeed, the growth is slower than $L\sim t^{1/2}$
up to times as long as $t\simeq 5\times 10^3$. From $t\simeq 5\times 10^3$ onwards, on the other hand,
finite size effects start to appear, as it can be seen from the fact that curves for different
system sizes start to separate and -- particularly -- because the growth becomes faster than
$L(t)\sim t^{1/2}$. These effects reduce the regime where finite-size effects are absent and scaling 
properties can be studied to a very narrow time range (this should compared to the
cases with $T_f=0$ and $T_f=1.5$, where such range is much wider).
As a consequence the data for $G$ and $C$ do not obey the scalings (\ref{scalG},\ref{scalC})
and, when trying to collapse them as in Figs. \ref{G_to_zero},\ref{C_to_zero},\ref{G_to_15},\ref{C_to_15}
the result is poor. This is seen for the autocorrelation function in Fig. \ref{C_to_22}
(the collapse looks better than what really is because the plot is compressed).
Notice however that the exponent $\lambda _c\simeq 1/8$ is quite clearly observed.
It is interesting to observe that, despite the presence of preasymptotic and finite-size effects,
the scaling behavior (\ref{scalChi}) of the response function 
is robust. Indeed, as it can be seen in Fig. \ref{Chi_to_22}, data can be collapsed reasonably well
by plotting $L(s)^{\alpha}\chi (t,s)$ against $L(t)/L(s)$. Due to the noisy character of the data
(due to large thermal fluctuations) a precise estimate of the exponent $\alpha $ is not possible.
However, the comparison between the two values $\alpha =5/8$ and $\alpha =0.56$ shows that the latter
provides a slightly better collapse. This is in agreement with the expectation that roughening 
effects (which, as discussed above, are associated to an exponent $\alpha =1/2$) are
more pronounced at high $T_f$.

\section{The large-N model}
\label{largen}

The dynamics of a classical magnetic system with a vectorial order-parameter with $N$ components 
can be exactly solved in the large-$N$ limit \cite{corbcastzan}. This provides an 
analytic framework to interpret the behavior of physical systems with finite-$N$ and to 
compute the scaling properties at a semi-quantitative level.
In this section, by solving the large-$N$ model for growth kinetics and computing correlation and response
functions we show how some of the features observed in the numerical simulations 
can be interpreted in this analytical framework. The present solution closely follows and 
generalizes the
one contained in \cite{corberi2002}, to which we generally refer the reader for specific details,
extending the results to the case of a quench from an initial critical state.

\subsection{Model definition}

We consider a $d$-dimensional magnetic system with a vectorial order parameter 
$\vec{\phi}=\left( \phi_1,...,\phi_N \right)$
and a Ginzburg-Landau Hamiltonian 
\begin{equation}
\mathcal{H}\left[\vec{\phi}\right] = \int_V d^dx \, \left[ \frac{1}{2}\left(\nabla\vec{\phi}\right)^2 + \frac{r}{2}\vec{\phi}^2 + \frac{g}{4N}\left(\vec{\phi}^2\right)^2 \right], \label{hamiltonian-general}
\end{equation}
where $V$ is the volume and $r$ and $g$ are constants ($r<0$, $g>0$). 

In the case of a non-conserved order parameter we are considering here the Langevin equation of motion 
reads
\begin{equation}
\frac{\partial \vec{\phi}\left(\vec{x},t\right)}{\partial t} = -\frac{\delta \mathcal{H}\left[\vec{\phi}\right]}{\delta \vec{\phi}\left(\vec{x},t\right) } + \vec{\eta}\left(\vec{x},t\right), \label{langevin}
\end{equation}
where $\vec \eta\left(\vec{x},t\right)$ is a Gaussian white noise with expectations
\begin{eqnarray}
\overline {\eta_\alpha \left(\vec{x}\right) } &=& 0 \\
\overline {\eta_\alpha\left(\vec{x}\right) \eta_\beta\left(\vec{x}'\right) } &=& 2T \delta_{\alpha\beta} \delta\left(\vec{x}-\vec{x}'\right) \delta\left( t-t' \right)
 \label{gaussian-noise}
\end{eqnarray}
Here $\eta _\alpha $ is the $\alpha$ component of the vector $\vec \eta $, $T$ is the temperature of the thermal bath, and $\overline {\cdots}$ denotes an average over thermal fluctuations, namely over
different realizations of $\vec \eta$. 

In the large-$N$ limit the substitution $\vec{\phi}^2(\vec x,t)\to \langle \vec \phi ^2 (\vec x,t)\rangle$, where
$\langle\cdot\rangle$ denotes an ensemble average, namely over thermal noise and initial conditions,
becomes exact. Notice that the quantity $S(t)=\frac{1}{N}\langle \vec \phi ^2 (\vec x,t)\rangle$ does not depend on 
$\vec x$ due to space homogeneity.
In terms of  the Fourier transform 
$\vec \phi \left(\vec{k},t\right) = \int_V d\vec x \, \vec{\phi}\left(\vec{x},t\right) \exp\left(i\vec{k}\cdot \vec{x}\right)$ of $\vec \phi (\vec x,t)$ the evolution (\ref{langevin}) then  reads
\begin{equation}
\frac{\partial \vec{\phi}\left(\vec{k},t\right)}{\partial t} = -\left[k^2 + I(t)\right]\vec{\phi}\left(\vec{k},t\right) + \vec{\eta}\left(\vec{k},t\right), \label{foutrans-eq-motion}
\end{equation}
where $I(t)$ is the self-consistent function
\begin{equation}
I(t)= r + gS(t). 
\label{self-consist-func}
\end{equation}

\subsection{Non-equilibrium dynamics}

The general solution of the formally linear Eq.~(\ref{foutrans-eq-motion}) is
\begin{equation}
\phi(\vec{k},t)={\cal R}(k,t,0)\,\phi (\vec{k},0)+\int\limits_0^t d t' {\cal R}(k,t,t')\,\eta(\vec{k},t'), \label{generalsolution}
\end{equation}
where we denote by $\phi $ one of the equivalent components 
$\vec \phi _\alpha $ of the order parameter  and the response function ${\cal R}$,
which only depends on the modulus $k=\vert \vec k \vert$ of the wave-vector, is given by
\begin{equation}
 {\cal R}(k,t,t') = \frac{Y(t')}{Y(t)} e^{-k^2 (t-t')}, \label{response-function}
\end{equation}
with $Y(t) = \exp\left[ \int_0^t d s I(s) \right]$, $I(t)$ being the self-consistent function defined 
in~(\ref{self-consist-func}). The squared quantity $Y^2(t)$ obeys the following differential equation
\begin{equation}
\frac{d Y^2(t)}{d t} = 2 I(t) Y^2(t). \label{diff-eq-selfconsisfac}
\end{equation}
This equation contains, hidden in $I(t)$, the unknown $S(t)=\langle \phi^2 \rangle $ which is related to
the structure factor (the Fourier transform of the equal time correlation function $G(r,t)$)
\be
{\cal C}(k,t)=\langle \phi (\vec k,t) \phi (-\vec k,t)\rangle
\label{strucfac}
\ee
by
\be
\langle \phi^2 \left( \vec{x},t \right) \rangle =\frac{1}{(2\pi)^d}\int d\vec k \, {\cal C}(k,t)
e^{-\frac{k^2}{\Lambda ^2}},
\ee
where the smooth cut-off around $|\vec k| \sim \Lambda$ mimics the presence of 
a lattice and regularizes the theory in the ultraviolet sector, and $d$ is the spatial dimension. 
Using Eq. (\ref{generalsolution}) to build products of order parameter fields and
plugging them into Eq. (\ref{strucfac}), using the expectations of the noise (\ref{gaussian-noise})
one arrives at the following expression 
\begin{equation}
{\cal C}(k,t) = {\cal R}^2\left(\vec{k},t,0\right) {\cal C}(k,0) + 2 T \int\limits_0^t d t' {\cal R}^2 \left(\vec{k},t,t'\right). \label{strfac-respfunc}
\end{equation}

We now specify the initial configuration of the order parameter as
\begin{eqnarray}
\langle \phi(\vec{k},0)\rangle &=& 0 \nonumber \\
\langle \phi(\vec{k},0)\phi(\vec{k}',0)\rangle &=& (2\pi)^d 
\frac{\Delta}{k^\mu} \delta(\vec{k}+\vec{k}')
\label{initial condition}
\end{eqnarray}
This form contains, as special cases, an initial equilibrium state at $T_i=\infty$,
with the choice $\mu =0$, and that of a critical state at $T_c$
for $\mu = 2$, since in the large-$N$ model the critical exponent $\eta $ is 
$\eta =0$ \cite{corberi2002}. The solution presented below, however, 
is valid also for correlated initial states with different values of $\mu $.
We consider a quench from such an initial condition to
the final temperature $T_f$.
Using the above initial conditions in Eq. (\ref{strfac-respfunc}) one has
\begin{equation}
{\cal C}(\vec{k},t) = {\cal R}^2\left(\vec{k},t,0\right) \frac{\Delta}{k^\mu} + 2 T_f \int\limits_0^t d t' {\cal R}^2 \left(\vec{k},t,t'\right) 
\label{strfac-respfunc2}
\end{equation}
and inserting this form into Eq. (\ref{diff-eq-selfconsisfac}) one is left with the following
integro-differential equation
\begin{equation}
\frac{d Y^2(t)}{d t} = 2 r Y^2(t) + 2 g \Delta f\left(t+\frac{1}{2 \Lambda^2} ; \mu \right) + 4 g T_f \int\limits_0^t d t' f\left( t-t'+\frac{1}{2 \Lambda^2} ; 0 \right) Y^2(t) \label{diff-eq-ysqr}
\end{equation}
where
\begin{equation}
f(x;\mu)=\frac{1}{(2\pi)^d}\int d \vec k \,k^{-\mu} e^{ -2k^2 x } = 2^{-\frac{3d-\mu}{2}}\pi^{-\frac{d}{2}} x^{-\frac{d-\mu}{2}} \frac{\Gamma\left( \frac{d-\mu}{2}  \right)}{\Gamma\left(  \frac{d}{2} \right)} \label{g-f-definitions}
\end{equation}
and $\Gamma $ is the Euler special function.

Eq. (\ref{diff-eq-ysqr}) is a closed equation for the quantity $Y$. Its solution will be discussed in Appendix \ref{appA},\ref{appB}. As we will show in Sec. \ref{largeNcorr}, any observable can be expressed in terms of $Y$.

\subsection{Observables} \label{largeNcorr} \label{secstrfac-respfunc}

From the knowledge of $Y(t)$ it is possible to compute the two-time quantities considered
in this paper. Extending the definition (\ref{strucfac}) to a two-time correlation
\be
{\cal C}(k,t,s)=\langle \phi (\vec k,t) \phi (-\vec k,s)\rangle
\label{strucfac2}
\ee
and using the expression (\ref{generalsolution}) to build the $\phi$-product one arrives at
\begin{equation}
{\cal C}\left(\vec{k},t,s\right) = {\cal R}\left(\vec{k},t,0\right) {\cal R}\left(\vec{k},s,0\right) \frac{\Delta}{ k^{\mu}} + 2T_f \int\limits_0^{s} dt' {\cal R}\left(\vec{k},t,t'\right) {\cal R}\left(\vec{k},s,t'\right), 
\label{two-time-structure-factor2}
\end{equation}
from which the autocorrelation function is readily obtained as
\be
C(t,s)=\frac{1}{(2\pi)^d}\int d\vec k\, 
{\cal C}(k,t,s)e^{-\frac{k^2}{\Lambda ^2}}.
\ee
Using the expression (\ref{response-function}) one arrives at
\be
C(t,s)=\frac{1}{Y(t)Y(s)}\left [ f \left (\frac{t+s}{2}+\frac{1}{2\Lambda ^2};\mu\right ) \Delta
+2T_f \int _0 ^s dt' f\left (\frac{t+s}{2}-t'+\frac{1}{2\Lambda ^2};0\right ) Y^2(t') \right ].
\ee
The first term on the r.h.s. is responsible for the aging properties \cite{corberi2002}. Using 
the expression (\ref{g-f-definitions}) for $f$ and the expressions for $Y$ derived
in Appendix \ref{appA}, focusing on the large time sector 
$t+s\gg \frac{1}{\Lambda ^2}$ one has
\be
C(t,s)=M^2 \left [\frac{4x}{x^2+1}\right ]^{\frac{(d-\mu)}{2}}
\ee
where $M^2=-\frac{r}{g} \frac{T_c-T_f}{T_c}$ is a constant
[the critical temperature of the large-$N$ model is 
$T_c=\frac{-r (4\pi)^{d/2}}{2g\Lambda ^{d-2}}(d-2)$], 
$x=L(t)/L(s)$ and $L(t)\sim t^{1/2}$. 
This result shows that the autocorrelation function takes the general scaling form of 
Eq. (\ref{scalC}) and that
\be
\lambda _C=\frac{d-\mu}{2}.
\ee
Therefore, there is a memory of the initial condition -- through the value of $\mu $ --
in the exponent $\lambda _C$. Notice that going from $T_i=\infty$ (i.e. $\mu =0$)
to $T_i=T_c$ (i.e. $\mu =2$) reduces the autocorrelation exponent, as it also true in the
Ising model (see Sec. \ref{SIMUL}). The actual value of this exponent in the 
large-$N$ model is different from the one observed in the scalar case, 
as expected.

Let us now consider the response function. It is easy to show \cite{corberi2002} that 
 the impulsive auto-response $R(t,t')$ defined in Eq. (\ref{defR}) is related to
 ${\cal R}(k,t,t')$ by
 \be
R(t,t')=\frac{1}{(2\pi )^d}\int d\vec k \,{\cal R}(k,t,t')\,e^{-\frac{k^2}{\Lambda ^2}}.
\ee
Using the expressions (\ref{response-function},\ref{g-f-definitions}) and the behavior
of $Y(t)$ derived in Appendix \ref{appA} to compute this
quantity and plugging the result in the definition (\ref{defChi}) of the integrated response
(the quantity measured in the numerical simulations of Sec. \ref{SIMUL}) one 
obtains, for large $s$, the general scaling form (\ref{scalChi}) with 
\be
\alpha = d-2
\label{alfaninf}
\ee 
and
\be
h(x)=(4\pi )^{-\frac{d}{2}}x^{2-d}\int _{x^{-2}}^1 dz \,z^{-\frac{d-\mu}{4}}\,
(1-z)^{-\frac{d}{2}}.
\label{scalhninf}
\ee
Eq. (\ref{alfaninf}) shows that the response exponent $\alpha$ is independent
of $\mu $, therefore it is not touched by changing the initial condition.
This is indeed what we found in Sec. \ref{SIMUL} also in the simulations
of the Ising model. Notice that for $2<d\le 4+\mu$ (a range where all physically
relevant cases are included), since the integral in Eq. (\ref{scalhninf}) converges,
one finds the general behavior (\ref{largeh}) and the constraint (\ref{lameqalf}),
so that also $\lambda _\chi$ does not change with $T_i$. The shape of the scaling
function $h$, instead, changes. In particular, the effect of raising $\mu$ is to lower $h(x)$,
as indeed it was found also in the Ising model (see Sec. \ref{SIMUL}) since the response function
is smaller with $T_i=T_c$ than with $T_i=\infty$. 

In the large-$N$ model the different sensitivity of the correlation and of the response function 
exponents have a clear mathematical origin.
We have seen that the initial condition plays a role in determining the
time-behavior of the self-consistent quantity $Y$, Eq. (\ref{mainresy}).
Eq. (\ref{response-function}) shows that this different time-behavior is the only 
effect of the initial condition on the wave-vector-resolved response function ${\cal R}(k,t,s)$.
The situation is different for the autocorrelation function (\ref{strucfac2}).
Indeed, Eq. (\ref{two-time-structure-factor2}) shows that the different spatial organization
of the correlation is explicitly determined by the one in the
initial state through the factor $\frac{\Delta }{k^\mu}$. This makes the effect of a different
initial condition much more pronounced than in the response function, producing a different
exponent $\lambda _C$. It can be observed that the role of the factor $\frac{\Delta}{k^\mu}$ in
Eq. (\ref{two-time-structure-factor2}) is such to weight more the contribution of
the small wave-vectors if the initial condition is critical than in the case of a disordered ones.
We have already noticed in Sec. \ref{SIMUL} that in the scalar case 
the different behavior of the system at
large distances, where memory of the initial condition is retained, might be the origin of the different value
of the exponent $\lambda _C$ in the quenches from $T_i=\infty$ or from $T_i=T_c$.
We see here that a similar property is shared by the analytically tractable large-$N$ model.

\section{Summary and conclusions} \label{concl}

In this paper we have discussed the results of a rather general investigation of the
phase-ordering process observed in a ferromagnetic system described by the 
Ising model with Glauber single-spin flip dynamics quenched from equilibrium
states at infinite temperature or at the critical one $T_c$. We have considered three values of
the final quench temperature $T_f$ in order to scan the region  $0\le T_f <T_c$.
When a deep quench is made from an infinite initial temperature $T_i$ to a vanishing
one $T_f=0$, all the quantities considered show an excellent agreement with the expected
dynamical scaling forms. Similar results are also found in quenches from
the critical state, but severe finite-size effects restrict the region where scaling is observed
to a much smaller time/space region than in the quench from $T_i=\infty$, as an effect 
of the correlated initial state. Upon raising $T_f$ (for any choice of the initial state),
the quality of the data collapse predicted by dynamical scaling gets progressively poorer
until, in a shallow quench at $T_f=2.2$, scaling is basically lost. This is due partly to the 
stationary term of Eq. (\ref{calG}) becoming more important and to 
the relevance of pre-asymptotic effects due to the proximity of the critical point. 

Besides providing a comparative study of the effects of changing the initial and the final
quench temperature, our study includes the first determination of the response function 
in quenches from $T_i=T_c$ and in those to $T_f=0$. We have shown that, while 
starting from $T_i=T_c$ instead of $T_i=\infty$ 
does change the aging properties of the process, 
as witnessed by the markedly different behavior of the autocorrelation function,
the universal properties of the response function are basically insensitive to the different initial conditions.
The same effect is found in the behavior of the exactly solvable large-$N$ model
and has been interpreted as due to the large sensitivity of the autocorrelation   
function to the large-scale properties of the system, which are reminiscent of the 
correlated initial configuration, whereas the largest contributions to the response
are provided at small scales.

The computation of the dynamical susceptibility in a quench to a vanishing final temperature 
allows a rather accurate determination of the response function exponent which turns out
to be consistent with $\alpha=\frac{1}{2}\lambda _C=5/8$, at variance with previous 
determinations lying in the range $[0.5-0.56]$ obtained at 
finite $T_f$. This seems to rule out the
conjecture that the value $a=\frac{1}{2}$ is the asymptotic one at $T_f=0$.
Instead, our data at $T_f>0$ suggest that the smaller exponent in this case might be a pre-asymptotic
effect associated to the thermal roughening of interfaces.
The observed relation $\alpha = \frac{1}{2} \lambda _C$ has presently no physical
interpretation. 
We hope that the results of this paper will refresh the attention on the non-equilibrium 
response exponent possibly providing a thorough understanding.

\vspace{1cm}
{\bf Acknowledgments}

We thank Eugenio Lippiello and Marco Zannetti for discussions. 
We acknowledge financial support by MURST PRIN 2010HXAW77\_005.

\appendix

\section{Solution of the equation for $Y$} \label{appA}

\begin{figure}[t]
\rotatebox{0}{\resizebox{.45\textwidth}{!}{\includegraphics{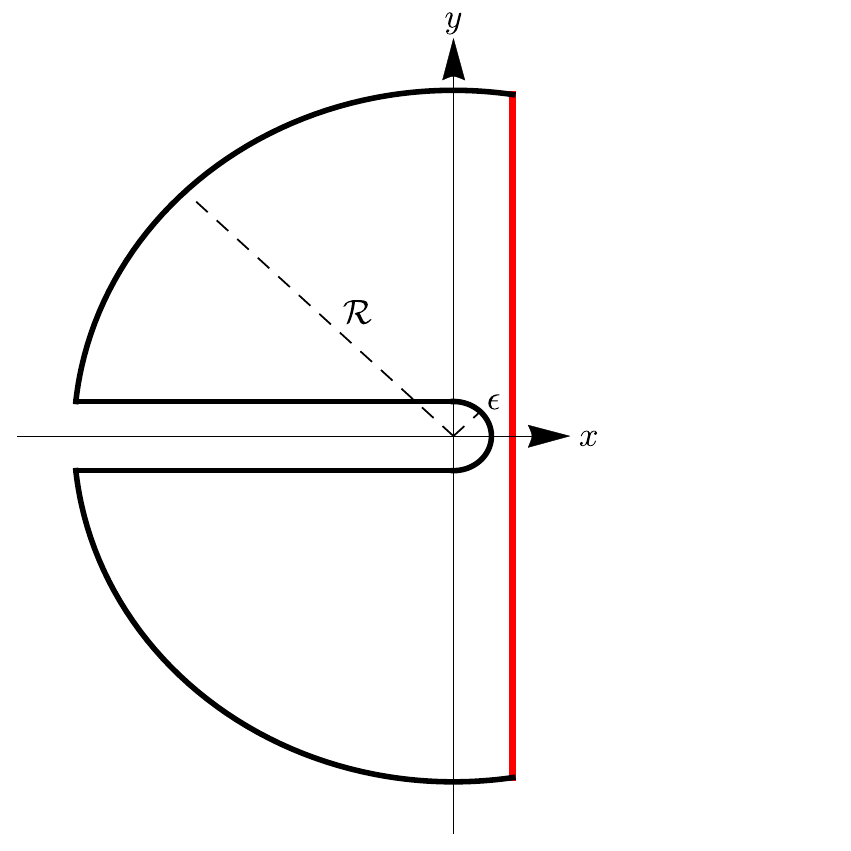}}}
	\caption{(Color on-line). Bromwich contour to calculate $Y^2(t)$. The definition of the inverse Laplace transform involves integrating over the (red) vertical segment of the contour.} \label{figure-bromwich-contour}
\end{figure}

In order to solve Eq.~(\ref{diff-eq-ysqr}), following \cite{braynewmann}, we Laplace transform
it to obtain
\begin{equation}
Y_L^2(s)=\frac{1+2g\Delta f_L(s;\mu)}{s-2r-4gT_f f_L(s;0)}, \label{ysqr-laplace}
\end{equation}
where $Y_L^2(s)$ is a shorthand for the Laplace transform of $Y^2(t)$, and
\begin{equation}
f_L(s;\mu)=2^{-\frac{3d-\mu}{2}}   s^{\frac{d-\mu}{2}-1}\,\frac{\Gamma\left(\frac{d-\mu}{2}\right)\Gamma\left(1-\frac{d-\mu}{2},\frac{s}{2\Lambda ^2}\right)}{\Gamma\left(\frac{d}{2}\right)}. \label{f-laplace}
\end{equation}
Here $\Gamma(a,y)$ is the (upper) incomplete gamma function. The final step is to calculate the inverse Laplace transform
\begin{equation}
Y^2(t)=\frac{1}{2\pi i}\int\limits_{\sigma-i\infty}^{\sigma+i\infty}ds \, e^{st} \,Y_L^2(s), \label{inverse-laplace}
\end{equation}
which is done, using standard techniques, by previously extending the integral in Eq. (\ref{inverse-laplace}) 
along the closed Bromwich contour $	\mathcal{B}$ shown in Fig.~\ref{figure-bromwich-contour}.

According to the complex inversion theorem $\mathcal B$ has to be chosen in such a way that
all the poles of  $Y_L^2$ lie on the left of the straight vertical line (marked red in 
Fig. \ref{figure-bromwich-contour}). In this case, a direct analysis of the denominator of $Y_L^2(s)$ 
in Eq. (\ref{ysqr-laplace}) shows \cite{braynewmann} that poles only exist when a quench is made to $T_f\ge T_c$,
so there is no restriction on the position of the vertical segment of $\mathcal B$ 
in the cases with $T_f< T_c$ we are interested here. We place it along the imaginary
axis, therefore. Notice that the contour has to be deformed as 
to avoid the branch cut along the negative real axis because the expression (\ref{f-laplace})
implies that $f_L$, and then also $Y_L$, contains fractional powers of $s$, in general.
We also remark that since cases where $d \leq 2$ lead to $T_c=0$, 
we shall focus on dimensions $d>2$.

Letting the radius ${\cal R}$ of the outer circle go to infinity and the one $\epsilon $ of the 
inner one vanish, the only non-vanishing contributions along the contour $\mathcal B$
come from the horizontal segments along the branch cut and from the vertical one. Then one has 
\begin{equation}
\int\limits_{\mathcal{B}} Y_L^2(s) = \int\limits_{\sigma-i\infty}^{\sigma+i\infty} ds \, e^{st} Y_L^2(s) + \int\limits_0^\infty dx \, e^{-xt} \left( Y_L^2(s) |_{s=xe^{i\pi}} - Y_L^2(s) |_{s=xe^{-i\pi}} \right) = 0, \label{bromwich-contour-integral}
\end{equation}
where the last equality follows from Cauchy residue theorem, since there are no singularities inside 
$\mathcal B$. Recalling Eq. (\ref{inverse-laplace}) one finds
\begin{equation}
Y^2(t) = -\frac{1}{2 \pi i} \left[ \int\limits_0^\infty dx \, e^{-xt} \left( Y_L^2(s) |_{s=xe^{i\pi}} - Y_L^2(s) |_{s=xe^{-i\pi}} \right) \right]. \label{inverse-laplace-2}
\end{equation}
Since we are interested in the behavior of the model in the asymptotic time domain, we can 
evaluate Eq. (\ref{inverse-laplace-2}) by expanding $Y(s)$ for small values of $s$,
which can be done in different ways according to the value of the exponents $d,\mu$ 
as it is explained below.   

\subsection{$\frac{d-\mu}{2}$ and $\frac{d}{2}$ non-integer quantities} \label{expodd}

When $\frac{d}{2}$ and $\frac{d-\mu}{2}$ are non-integer quantities the expansion of $Y^2_L$ for
small $s$ is given in Eq. (\ref{ysqr-expansion}) of the Appendix.
Then the anti-Laplace transform of $Y_L^2(s)$ is calculated by solving the right-hand side of~(\ref{inverse-laplace-2}) which leads to the integral
\begin{eqnarray}
Y^2(t) &= \frac{1}{ 2\pi iD_0} \int\limits_0^\infty dx\, e^{-xt} \left[ \left.  \left(N_0 + N_1 s^{\frac{d-\mu}{2}-1} - \frac{D_1 N_0}{D_0} s^{\frac{d}{2}-1} - \frac{N_1D_1}{D_0} s^{d-2-\frac{\mu}{2}} \right) \right\vert_{s=xe^{i\pi}} \right. \nonumber \\
&\qquad - \left. \left.  \left(N_0 + N_1 s^{\frac{d-\mu}{2}-1} - \frac{D_1 N_0}{D_0} s^{\frac{d}{2}-1} - \frac{N_1D_1}{D_0} s^{d-2-\frac{\mu}{2}} \right) \right\vert_{s=xe^{-i\pi}} \right]
, \label{anti-laplace-complete}
\end{eqnarray}
where the constants $D_0,D_1,N_0,N_1$ are given in Eq. (\ref{coeffs}).

The terms with an integer exponent on the right-hand side of Eq.~(\ref{anti-laplace-complete}) cancel out,
and the same happens for the real part of the terms with a non-integer exponent. Integrating the imaginary parts of the latter easily leads to
\begin{eqnarray}
Y^2(t) &=& \frac{1}{\pi D_0} \left\{ N_1 \Gamma\left(\frac{d-\mu}{2}\right) \sin\left[\pi\left(\frac{d-\mu}{2}-1\right)\right] t^{-\frac{d-\mu}{2}} - \frac{D_1 N_0}{D_0} \Gamma\left(\frac{d}{2}\right) \sin\left[\pi\left(\frac{d}{2}-1\right) \right] t^{-\frac{d}{2}} \right \} 
.\label{ysqr-full}
\end{eqnarray}
In short we have the result
\be
Y^2(t)\simeq A_{d,\mu}t^{-\frac{d-\mu}{2}}
\label{mainresy}
\ee
with 
$A_{d,\mu}=\frac{N_1}{\pi D_0} \Gamma\left(\frac{d-\mu}{2}\right) \sin\left(\pi\left(\frac{d-\mu}{2}-1\right)\right)$ for $\mu>0$ and $A_{d,0}=\frac{1}{\pi D_0^2} \Gamma\left(\frac{d}{2}\right) \sin\left(\pi\left(\frac{d}{2}-1\right)\right)[D_0N_1-D_1N_0]$ for $\mu =0$.

\subsection{$\frac{d-\mu}{2}$ and/or $\frac{d}{2}$ integer quantities}

When the exponents entering the function $f_L$ are (negative) integer quantities the small-$s$ expansion
of the incomplete gamma function leads to the expression (\ref{f-laplace-series-even}). 
Therefore, when computing the small-$s$ expansion of $Y^2$ through Eq. (\ref{ysqr-laplace}) different cases
must be considered, namely when $\frac{d}{2}$ or $\frac{d-\mu}{2}$ or both are integer.
Using the expansions of $f_L$ and $Y^2_L$ given in Appendix \ref{appB} and proceeding as in Sec. \ref{expodd} it is easy to show that 
the integral~(\ref{inverse-laplace-2}) gives
\begin{eqnarray}
Y^2(t) &=& \frac{\cos\left[\left( \frac{d-\mu}{2}-1 \right) \pi \right] M_1 \Gamma\left(\frac{d-\mu}{2}\right) }{D_0} t^{-\frac{d-\mu}{2}} + \frac{\sin\left[\left(\frac{d}{2}-1\right)\pi\right] M_0 D_1 \Gamma\left(\frac{d}{2}\right)}{\pi D_0^2} t^{-\frac{d}{2}} \label{main-result-even-odd} \nonumber \\
Y^2(t) &=& \frac{-\sin\left[\left( \frac{d-\mu}{2}-1 \right) \pi \right] N_1 \Gamma\left(\frac{d-\mu}{2}\right) }{\pi E_0} t^{-\frac{d-\mu}{2}} + \frac{\cos\left[\left(\frac{d}{2}-1\right)\pi\right] N_0 E_1 \Gamma\left(\frac{d}{2}\right)}{ E_0^2} t^{-\frac{d}{2}} \label{main-result-odd-even} \nonumber \\
Y^2(t) &=& \frac{-\cos\left[\left( \frac{d-\mu}{2}-1 \right) \pi \right] M_1 \Gamma\left(\frac{d-\mu}{2}\right) }{E_0} t^{-\frac{d-\mu}{2}} + \frac{\cos\left[\left(\frac{d}{2}-1\right)\pi\right] M_0 E_1 \Gamma\left(\frac{d}{2}\right)}{E_0^2} t^{-\frac{d}{2}} \label{main-result-even-even}.
\end{eqnarray}
for i) $d-\mu$ even and $d$ odd, ii) $d-\mu$ odd and $d$ even and iii) both $d-\mu$ and $d$ even,
respectively. In any case
one has the same result 
of Eq. (\ref{mainresy}), where the value of $A_{d,\mu}$ can be evinced from  
Eqs. (\ref{main-result-even-even}).

\section{Expansion of some functions in Laplace space} \label{appB}

When $d-\mu>0$ is odd we have
\begin{eqnarray}
f_L(s;\mu) &=&  \frac{\Lambda^{d-\mu-2} \Gamma\left(\frac{d-\mu}{2}\right)}{2^{d+1} \Gamma\left(\frac{d}{2}\right)} \left( \frac{s}{2\Lambda^2} \right)^{\frac{d-\mu}{2}-1} \pi^{-\frac{d}{2}} e^{\frac{s}{2\Lambda^2}} \left[ \Gamma\left(1-\frac{d-\mu}{2}\right) - \left(  \frac{s}{2\Lambda^2}\right)^{1-\frac{d-\mu}{2}} \sum\limits_{n=0}^{\infty} \frac{    \left( -\frac{s}{2\Lambda^2}   \right)^n }{ \left(n+1-\frac{d-\mu}{2}\right)n! } \right]  \nonumber \\
&=&-\frac{  (4\pi)^{-\frac{d}{2}} \left(\Lambda^2\right)^{ \frac{d-\mu}{2}-1 } \Gamma\left( \frac{d-\mu}{2} \right)}{\left( 2-\left(d-\mu\right) \right) \Gamma\left( \frac{d}{2} \right)} + \frac{(4\pi)^{-\frac{d}{2}} \Gamma\left( \frac{d-\mu}{2} \right) \Gamma\left( 1-\frac{d-\mu}{2} \right)}{ 2^{\frac{d-\mu}{2}} \Gamma\left( \frac{d}{2} \right)} s^{\frac{d-\mu}{2}-1} \nonumber \\
&&\quad - \frac{(4\pi)^{-\frac{d}{2}}\left( \Lambda^2 \right)^{\frac{d-\mu}{2}-2} \Gamma\left( \frac{d-\mu}{2} \right)}{\left((d-\mu)-4\right) \left((d-\mu)-2\right) \Gamma\left(\frac{d}{2}\right)} s + o(s^2).
\label{f-laplace-series-odd}
\end{eqnarray}
Plugging this expression into Eq.~(\ref{ysqr-laplace}) and
retaining the leading terms for $s\to 0$ one has 
\begin{equation}
Y_L^2(s) \simeq \left(-D_0\right)^{-1} \left(N_0 + N_1 s^{\frac{d-\mu}{2}-1} - \frac{D_1 N_0}{D_0} s^{\frac{d}{2}-1} - \frac{N_1D_1}{D_0} s^{d-2-\frac{\mu}{2}} \right)	\label{ysqr-expansion}
\end{equation}
where
\begin{eqnarray}
D_0 = 2r + 4gT_f \frac{\left(4\pi\right)^{-\frac{d}{2}} \Lambda^{d-2} }{ d-2 }&,&
D_1 = 4gT_f \left(8\pi\right)^{-\frac{d}{2}} \Gamma\left(1-\frac{d-\mu}{2}\right) , \nonumber \\
N_0 = 1 + 2g\Delta \frac{\left(4\pi\right)^{-\frac{d}{2}} \Lambda^{d-2-\mu} \Gamma\left(\frac{d-\mu}{2}\right)}{\left(d-2-\mu\right) \Gamma\left(\frac{d}{2}\right)}&,&
N_1 = 2g\Delta \frac{ \left(8\pi\right)^{-\frac{d}{2}} 2^{\frac{\mu}{2}} \Gamma\left(\frac{d-\mu}{2}\right) \Gamma\left(1-\frac{d-\mu}{2}\right) }{\Gamma\left(\frac{d}{2}\right)}.
\label{coeffs}
\end{eqnarray} 

Similarly, we can write the $f_L(s;\mu)$ as a power series when $d-\mu>0$ is even, which yields
\begin{eqnarray}
f_L(s;\mu) & = & \frac{\Lambda^{d-\mu-2} \Gamma\left(\frac{d-\mu}{2}\right) }{2^{d+1} \Gamma\left(\frac{d}{2}\right)} \left( \frac{s}{2\Lambda^2} \right)^{\frac{d-\mu}{2}-1} \pi^{-\frac{d}{2}} e^{\frac{s}{2\Lambda^2}} \left\{ \frac{(-1)^{\frac{d-\mu}{2}-1}}{\Gamma\left(\frac{d-\mu}{2} \right)} \left[ \psi\left(\frac{d-\mu}{2}\right)-\log\left(\frac{s}{2\Lambda^2}\right)  \right] \right. \nonumber \\
&\qquad& - \left. \left(\frac{s}{2\Lambda^2}\right)^{1-\frac{d-\mu}{2}} \sum\limits_{k=0; k\neq\frac{d-\mu}{2}-1}^{\infty}\frac{\left(-\frac{s}{2\Lambda^2}\right)^k}{\left(k+1-\frac{d-\mu}{2}\right) k!} \right\} \nonumber \\
& = & \frac{\left(\Lambda^2\right)^{\frac{d-\mu}{2}-1} \Gamma\left(\frac{d-\mu}{2}\right) \left( 1 - \delta_{0,\frac{d-\mu}{2}-1} \right)  }{\left(4\pi\right)^{\frac{d}{2}} \Gamma\left(\frac{d}{2}\right) \left(2-d+\mu\right)} -  \frac{\left(-1\right)^{\frac{d-\mu}{2}-1}   }{2^{\frac{d-\mu}{2}} \left(4\pi \right)^{\frac{d}{2}} \Gamma\left(\frac{d}{2}\right)} \log\left(\frac{s}{2\Lambda^2}\right)s^{\frac{d-\mu}{2}-1}  +  \frac{\left(-1\right)^{\frac{d-\mu}{2}-1} \psi\left(\frac{d-\mu}{2}\right)  }{2^{\frac{d-\mu}{2}} \left(4\pi \right)^{\frac{d}{2}} \Gamma\left(\frac{d}{2}\right)} s^{\frac{d-\mu}{2}-1} \nonumber \\
&\qquad & - \left( \frac{  1 - \delta_{1,\frac{d-\mu}{2}-1}  }{ 4-d+\mu } - \frac{ 1 - \delta_{0,\frac{d-\mu}{2}-1}  }{2-d+\mu} \right) \frac{\left(\Lambda^2\right)^{\frac{d-\mu}{2}-1} \Gamma\left(\frac{d-\mu}{2}\right) }{\left(4\pi\right)^{\frac{d}{2}} \Gamma\left(\frac{d}{2}\right) 2\Lambda^2} s +o(s)\label{f-laplace-series-even}
\end{eqnarray}
Here, the digamma function can be approximated for integer values by the series $\psi(n) = \sum_{k=0}^{n-1}k^{-1}-\gamma$, where $\gamma \simeq 0.5772$ is the Euler-Mascheroni constant. 

The series shown above can be used to expand~(\ref{ysqr-laplace}) which leads to three different cases,
namely when
\begin{itemize}
\item{i) $d-\mu$ even and $d$ odd}
\item{ii) $d-\mu$ odd and $d$ even}
\item{iii) $d-\mu$ even and $d$ even}
\end{itemize}
leading respectively to the following results
\begin{eqnarray}
Y_L^2(s) &=& D_0^{-1} \left( M_0 - M_1 \log\left(\frac{s}{2\Lambda^2}\right) s^{\frac{d-\mu}{2}-1} - \frac{M_0 D_1}{D_0} s^{\frac{d}{2}-1} \right) \label{d-nu_even-d_odd}\nonumber \\
Y_L^2(s) &=& E_0^{-1} \left( N_0 + N_1 s^{\frac{d-\mu}{2}-1} - \frac{N_0 E_1}{E_0} \log\left(\frac{s}{2\Lambda^2}\right) s^{\frac{d}{2}-1} \right) \label{d-nu_odd-d_even} \nonumber \\
Y_L^2(s) &=& E_0^{-1} \left( M_0 + M_1 \log\left(\frac{s}{2\Lambda^2}\right) s^{\frac{d-\mu}{2}-1} - \frac{M_0 E_1}{E_0} \log\left(\frac{s}{2\Lambda^2}\right) s^{\frac{d}{2}-1} \right) \label{d-nu_even-d_even}
\end{eqnarray}
The coefficients $D_i$ and $N_i$ are given in~(\ref{coeffs}) and
\begin{eqnarray}
M_0 &=& 1 + 2g\Delta \frac{\left(-1\right)^{\frac{d-\mu}{2}-1} \psi\left(\frac{d-\mu}{2}\right)  }{2^{\frac{d-\mu}{2}} \left(4\pi \right)^{\frac{d}{2}} \Gamma\left(\frac{d}{2}\right)} \delta_{0,\frac{d-\mu}{2}-1} - 2g\Delta \frac{\left(\Lambda^2\right)^{\frac{d-\mu}{2}-1} \Gamma\left(\frac{d-\mu}{2}\right) }{\left(4\pi\right)^{\frac{d}{2}} \Gamma\left(\frac{d}{2}\right) \left(2-d+\mu\right)} \left( 1 - \delta_{0,\frac{d-\mu}{2}-1} \right) \nonumber \\
M_1 &=& - 2g\Delta  \frac{\left(-1\right)^{\frac{d-\mu}{2}-1}   }{2^{\frac{d-\mu}{2}} \left(4\pi \right)^{\frac{d}{2}} \Gamma\left(\frac{d}{2}\right)} \nonumber \\
E_0 &=& -2r - 4gT_f \frac{\left(\Lambda^2\right)^{\frac{d}{2}-1} \left( 1 - \delta_{0,\frac{d}{2}-1} \right)  }{  \left(4\pi\right)^{\frac{d}{2}} \left(2-d\right)} \nonumber \\
E_1 &=& 4gT_f  \frac{\left(-1\right)^{\frac{d}{2}-1}   }{ \left(8\pi \right)^{\frac{d}{2}} \Gamma\left(\frac{d}{2}\right)}.
\end{eqnarray}

\end{document}